\documentclass[twocolumn,showpacs,prd,floatfix,axodraw]{revtex4}
\usepackage{mathrsfs}
\usepackage{graphicx,,booktabs,bm}
\usepackage{overpic}
\usepackage{color}
\usepackage{amssymb}

\usepackage{mathrsfs,bm,amsmath,amssymb}
\usepackage{longtable,lscape}
\usepackage{txfonts}
\usepackage{amssymb}
\usepackage{indentfirst}
\usepackage{graphicx,,booktabs}
\usepackage{multirow,ulem}

\usepackage{color}
\usepackage{amssymb}

\definecolor{cover}{rgb}{0.77,0.87,0.88}
\definecolor{blueone}{rgb}{0.1,0.1,.7}
\definecolor{citec}{rgb}{0.14,0.47,0.09}
\definecolor{two}{rgb}{0.0,0.5,0.}
\definecolor{three}{rgb}{.5,.1,0.15}
\usepackage[bookmarks=true,bookmarksopen=false,plainpages=false,breaklinks=true,
   bookmarksnumbered=true,hypertexnames=false,
   filecolor=blue,urlcolor=three,menucolor=three,
   linkcolor=three,citecolor=blueone, colorlinks,
   anchorcolor=blue,runcolor=pink,frenchlinks=red
   pdfstartview=FitH,pdftitle=title,%
   pdfauthor=author]{hyperref}

\begin{document}
\title{Production of the newly observed $\bar{T}_{c\bar{s}0}$ by kaon-induced reactions on a proton/neutron target}

\author{Yin Huang} \email{huangy2019@swjtu.edu.cn}
\affiliation{School of Physical Science and Technology, Southwest Jiaotong University, Chengdu 610031,China}

\author{Hao Hei}
\affiliation{School of Physical Science and Technology, Southwest Jiaotong University, Chengdu 610031,China}

\author{Jing-wen Feng}
\affiliation{School of Physics and Electronic Engineering, Sichuan Normal University, Chengdu 610101, China}

\author{Xurong Chen}
\affiliation{Institute of Modern Physics, Chinese Academy of Sciences, Lanzhou 730000, China}
\affiliation{School of Nuclear Science and Technology, University of Chinese Academy of Sciences, Beijing 100049, China}

\author{Rong Wang}
\affiliation{Institute of Modern Physics, Chinese Academy of Sciences, Lanzhou 730000, China}

\begin{abstract}
Recently, a new doubly charged tetraquark $T^{a}_{c\bar{s}0}(2900)^{++}$ and its neutral partner $T^{a}_{c\bar{s}0}(2900)^0$ at the  invariant mass
spectrum of $\pi{}D_s$ were observed by the LHCb Collaboration.  According to its properties, such as the mass and decay width, the $T^{a}_{c\bar{s}0}(2900)^{++/0}$
have been suggested to be a compact multi-quark state or a hadron molecule. In order to distinguish the various interpretations of the $T^{a}_{c\bar{s}0}(2900)^{++/0}$,
we investigate the possibility to study the $\bar{T}^{a}_{c\bar{s}0}(2900)$ [the antiparticle of $T^{a}_{c\bar{s}0}(2900)$] by kaon-induced reactions on a proton target
in an effective Lagrangian approach. The production mechanism is characterized by the $t$-channel $D$ meson exchange.  Our theoretical approach is based on the assumption
that $\bar{T}^{a}_{c\bar{s}0}(2900)$ can be either a $K^{*}D^{*}-D_s^{*}\rho$ molecule or a compact tetraquark state.  Using the coupling constants of the $\bar{T}^{a}_{c\bar{s}0}(2900)$ to $KD$ channel obtained from molecule or compact tetraquark picture of the $\bar{T}^{a}_{c\bar{s}0}(2900)$, we compute the cross-sections
for the process $K^{-}n\to{}\bar{T}^{a}_{c\bar{s}0}(2900)^{--}\Lambda^{+}_c$. The $\bar{K}N$ initial state interaction mediated by Pomeron and Reggeon exchanges is also
included, which reduces the production of the $\bar{T}^{a}_{c\bar{s}0}(2900)$.  Our calculations show that whether $\bar{T}^{a}_{c\bar{s}0}(2900)$ is a molecule or a compact tetraquark state, the cross-sections for the $K^{-}n\to{}\bar{T}^{a}_{c\bar{s}0}(2900)^{--}\Lambda^{+}_c$ reaction are of similar magnitude, ranging from approximately 0.075
nb to 0.270 nb.  However,a clearer comparison can be made by computing the cross-section of the $K^{-}n\to{}\bar{T}^{a}_{c\bar{s}0}(2900)^{--}\Lambda^{+}_c\to{}\pi^{-}D^{-}_s\Lambda^{+}_c$ reaction.  The results indicates that the cross-section for the molecule
assignment of $\bar{T}^{a}_{c\bar{s}0}(2900)$ can reach up to $1.83\times{}10^{-3}$ nb, which is significantly smaller than that of 0.122 nb by assuming $\bar{T}^{a}_{c\bar{s}0}(2900)$ as a compact tetraquark state.  Those results can be measured in future experiments, and can be used to test the nature of the $\bar{T}^{a}_{c\bar{s}0}(2900)$. Last, we also propose to search for the unreported charged tetraquark $T^{a}_{c\bar{s}0}(2900)^{+}$ in the $K^{-}p\to{}\bar{T}^{a}_{c\bar{s}0}(2900)^{-}\Lambda^{+}_c$ reaction.
\end{abstract}

\date{\today}


\maketitle
\section{Introduction}\label{sec:intro}
Studying hadrons with more complex internal structures than quark states, where mesons are composed of quark-antiquark pairs~\cite{Godfrey:1985xj},
and baryons are constructed from three quarks~\cite{Koniuk:1979vy, Isgur:1978wd}, is a prominent topic in particle physics.  We call them exotic
hadrons.  For an extended period, little noteworthy advancement has been made in the exploration of exotic states.  Only a few phenomena
suggesting that quarks $u/d/s$ can form exotic hadrons.  For example, In constituent quark models, the mass of the strange quark is approximately
50\% heavier than that of the $u/d$ quarks. This leads to questions regarding why the
$\Lambda(1405)$ has a significantly lower mass than the $N(1535)$. Moreover, it is puzzling to observe that the $N(1440)$, as a $N=2$ baryon, is much
lighter than the nucleon resonance $N(1535)$ with $N=1$, where $N$ is the main quantum number. These issues led Zou and his collaborators to propose
the existence of significant five-quark components in the nucleon and its resonances~\cite{Zou:2005xy,Riska:2005bh,An:2005cj}.  The mass inversion
problem could be easily understood if there are substantial five-quark $uuds\bar{s}$ components in the $N(1535)$~\cite{Zou:2010tc,Zou:2009zz}.
Furthermore, the five-quark configurations also provide a natural explanation for its large strange decay~\cite{Liu:2005pm,Xie:2007qt}.

However, it was in 2003 when the LHCb collaboration observed the $X(3872)$ in the $\pi^{+}\pi^{-}J/\psi$ mass spectra~\cite{Belle:2003nnu}, that this
field entered a new era.  From its observed decay mode, $X(3872)$ is known to consist of a pair of hidden-charm quarks and two pairs of light quarks.
Subsequent discoveries of several hidden-charm pentaquark states, including $P_c(4380)$, $P_c(4440)$, $P_c(4457)$, $P_c(4312)$, $P_{cs}(4338)$, and
$P_{cs}(4459)$, have further strengthened the belief in significant progress in the research of exotic hadrons~\cite{LHCb:2015yax, LHCb:2016ztz, LHCb:2016lve, LHCb:2019kea, LHCb:2020jpq, LHCb:2022ogu}.  The recent observation of a doubly charged tetraquark and its neutral partner by the LHCb collaboration in the analysis of the $B^0\to\bar{D}^{0}D^{+}_s\pi^{-}$ and $B^{-}\to\bar{D}^{-}D^{+}_s\pi^{-}$ reactions~\cite{LHCb:2022sfr} marks an another significant advancement in
the study of exotic hadrons.  Their masses and widthes were measured to be
\begin{align}
T^{a}_{c\bar{s}0}(2900)^0:~~~M&=2.892\pm{}0.014\pm{}0.015~~~~~{\rm GeV},\nonumber\\
                            \Gamma&=0.119\pm0.026\pm0.013~~~~~ {\rm GeV},\nonumber\\
T^{a}_{c\bar{s}0}(2900)^{++}:~~~M&=2.921\pm{}0.017\pm{}0.020~~~~~{\rm GeV},\nonumber
\end{align}
\begin{align}
                            \Gamma&=0.137\pm0.032\pm0.017~~~~~ {\rm GeV},
\end{align}
respectively.  Supposing the states belong to the same isospin triplet, the experiment also gave the shared mass and width,
\begin{align}
T_{c\bar{s}0}:~~~M&=2908\pm{}23~~~~~{\rm MeV},\nonumber\\
                            \Gamma&=136\pm25~~~~~~ {\rm MeV},
\end{align}

Similar to the challenges faced with other exotic hadrons, the true internal structure of these two newly observed mesons cannot be completely determined based on
existing experimental data.  Due to the close proximity of the mass of $T^{a}_{c\bar{s}0}(2900)$ to the $D^{}K^{}$ threshold, the authors of Ref.\cite{Chen:2022svh}
proposed a novel interpretation for the recently discovered $T^{a}_{c\bar{s}0}(2900)$. They suggest that these two states could be an isovector $D^{*}K^{*}$ molecular state
with quantum numbers $I(J^P)=1(0^{+})$. In an independent study conducted by the authors of Ref.\cite{Yue:2022mnf}, it was found that if the $T^a_{c\bar{s}0}(2900)^0$ is
indeed a molecular state formed by $D^{*0}K^{*0}$, its primary decay mode would likely be into $D^0K^0$, rather than the observed experimental decay channel $D_s^{+}\pi^{-}$.
The authors in Ref.~\cite{Agaev:2022eyk} support the $T^{a}_{c\bar{s}0}(2900)$ as a $D^{*}K^{*}$ molecule based on the analysis of the mass spectrum and
the partial widths using the QCD light-cone sum rule approach and soft-meson approximation.   The mass of the $T^a_{c\bar{s}0}$ was studied in the coupled-channel
approach and it was shown that $T_{c\bar{s}0}$ might be $D^{*}K^{*}-D_s^{*}\rho$ couple molecular state~\cite{Duan:2023lcj}.  However, Ref.~\cite{Ke:2022ocs} reached
a strikingly different conclusion, arguing that $T^a_{c\bar{s}0}$ should ont be considered as a $D^{*}K^{*}$ bound state, but instead might be compact tetraquarks.

Indeed, the newly observed two mesons can be assigned to be the  lowest $1S$ -wave tetraquark states~\cite{Liu:2022hbk} within the framework of a nonrelativistic
potential quark model. Their analysis indicates that the dominant decay mode is the $D_s^{*}\rho$.  Furthermore, the compact tetraquark explanation of $T^a_{c\bar{s}0}$
are also supported by estimates obtained from the multiquark color flux-tube model~\cite{Wei:2022wtr}.  QCD sum rules, informed by the examination of mass spectrum
and the two-body strong decays, have lead Refs.~\cite{Yang:2023evp,Lian:2023cgs} to classify $T^a_{c\bar{s}0}$ as compact tetraquark states.  And the studies also reveal
that the primary decay models for $T^a_{c\bar{s}0}$ involve $D_s\pi$ and $DK$ channels~\cite{Lian:2023cgs}.  The compact tetraquark candidates for $T^a_{c\bar{s}0}$
gains additional support from the Refs.~\cite{Jiang:2023rcn,Dmitrasinovic:2023eei,Ortega:2023azl}. We note that the threshold effect from the interaction between the
$D^{*}K^{*}$ and $D_s^{*}\rho$ channels and the kinetic effect from a triangle singularity for $T^a_{c\bar{s}0}$ are also proposed in Ref.~\cite{Molina:2022jcd} and
Ref.~\cite{Ge:2022dsp}, respectively.

In addition to analyzing the mass spectrum and decay width, exploring the production mechanism provides a more effective approach to evaluating the nature of
$T^a_{c\bar{s}0}$.  This is mainly due to the production process strong dependence on the internal structure of $T^a_{c\bar{s}0}$.  We find that whether $T^a_{c\bar{s}0}$
is a molecular state or a compact multi-quark state, it exhibits a significant $KD$ decay width.  This motivates our quest to search for $\bar{T}^{a}_{c\bar{s}0}(2900)^{--}$
in the $K^{-}n\to{}\bar{T}^{a}_{c\bar{s}0}(2900)^{--}\Lambda^{+}_c$ and $K^{-}n\to{}\bar{T}^{a}_{c\bar{s}0}(2900)^{--}\Lambda^{+}_c\to{}\pi^{-}D^{-}_s\Lambda^{+}_c$ reactions.  Notably, high-energy kaon beams are available at OKA@U-70~\cite{Obraztsov:2016lhp}, SPS@CERN~\cite{Velghe:2016jjw}, CERN/AMBER~\cite{Quintans:2022utc}, and potential upgrades
to the J-PARC kaon beam, enabling us to reach the necessary energy range for $\bar{T}^{a}_{c\bar{s}0}(2900)^{--}$ production~\cite{Nagae:2008zz}.  Consequently, the searching
for $\bar{T}^{a}_{c\bar{s}0}(2900)^{--}$ in the $K^{-}n\to{}\bar{T}^{a}_{c\bar{s}0}(2900)^{--}\Lambda^{+}_c$ and $K^{-}n\to{}\bar{T}^{a}_{c\bar{s}0}(2900)^{--}\Lambda^{+}_c\to{}\pi^{-}D^{-}_s\Lambda^{+}_c$
reactions becomes feasible.  This approach facilitates a straightforward differentiation between molecular and compact tetraquark states through the production process.

In this study, we examine the recently observed $\bar{T}^{a}_{c\bar{s}0}(2900)$ production in the $K^{-}n\to{}\bar{T}^{a}_{c\bar{s}0}(2900)^{--}\Lambda^{+}_c$ and $K^{-}n\to{}\bar{T}^{a}_{c\bar{s}0}(2900)^{--}\Lambda^{+}_c\to{}\pi^{-}D^{-}_s\Lambda^{+}_c$ reactions by considering $T^{a}_{c\bar{s}0}$ as a molecular state and a compact multiquark state, respectively.  A conclusive determination of the inner structure of $T^a_{c\bar{s}0}$ can be attained by comparing the obtained cross-section with future experimental data.  To enhance the reliability of our predictions, the effect from the $\bar{K}N$ initial state interaction (ISI) must be taken into account due to there exist
plenty of experimental information about the $\bar{K}N$ elastic interaction in the considering energy region. Moreover, we also proposed to search for its isospin partner $\bar{T}^{a}_{c\bar{s}0}(2900)^{-}$ in the $K^{-}p\to\bar{T}^{a}_{c\bar{s}0}(2900)^{-}\Lambda^{+}_c$ reaction.  This paper is organized as follows.  In Sec.~\ref{Sec: formulism},
we will present the theoretical  formalism.  In Sec.~\ref{Sec: results}, the numerical result will be given, followed by discussions and conclusions in the last section.

\section{FORMALISM AND INGREDIENTS}\label{Sec: formulism}
\begin{figure}[h!]
\begin{center}
\includegraphics[bb=90 600 950 720, clip, scale=0.6]{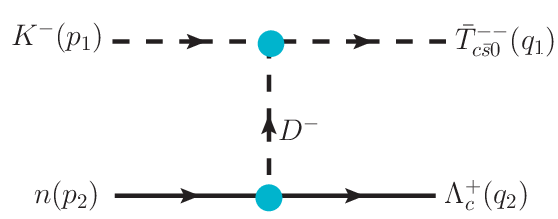}
\caption{Feynman diagram for the $K^{-}n\to{}\bar{T}^{a}_{c\bar{s}0}(2900)^{--}\Lambda^{+}_c$ reaction.  The contributions from the $t$-channel $D^{-}$ meson
exchange. We also show the definition of the kinematics ($p_1,p_2,q_1,q_2$) that we use in the present calculation.}\label{cc1}
\end{center}
\end{figure}

In this study, we investigate the feasibility of measuring $\bar{T}^{a}_{c\bar{s}0}(2900)^{--}$ in the $K^{-}n\to{}\bar{T}^{a}_{c\bar{s}0}(2900)^{--}\Lambda^{+}_c$
reaction.  We consider two scenarios for $T^a_{c\bar{s}0}$: one as a molecular state and the other as a compact multi-quark state.  The considered Feynman diagrams
are illustrated in Fig.~(\ref{cc1}), which includes only the $t$-channel $D^{-}$ meson exchange diagram.  And the ISI is represented by the red circle.   This
production process differs from the complex proton-proton collisions~\cite{LHCb:2022sfr}, as $\bar{T}^{a}_{c\bar{s}0}(2900)^{--}$ production in $K^{-}n\to{}\bar{T}^{a}_{c\bar{s}0}(2900)^{--}\Lambda^{+}_c$ occurs more simply.  The reason lies in the significantly lower required center-of-mass energies
compared to proton-proton collisions.  At these lower energies, we can neglect contributions from the $s$- and $u$-channels, which involve the creation of an
additional $c\bar{c}$ quark pair in kaon-induced production and are typically strongly suppressed.  Hence, the $K^{-}n\to{}\bar{T}^{a}_{c\bar{s}0}(2900)^{--}\Lambda^{+}_c$
reaction is expected to be primarily governed by Born terms through the $t$-channel $D^{-}$ exchanges, resulting in minimal background interference.

To calculate the diagrams depicted in Fig.~(\ref{cc1}), it is necessary to determine the effective Lagrangian densities corresponding to the relevant interaction
vertices.  In the case of $\Lambda_cND$ coupling, we adopt the Lagrangian densities employed in Ref.\cite{Dong:2014ksa,Zhu:2019vnr}
\begin{align}
{\cal{L}}_{\Lambda_cND} &= ig_{\Lambda_cND}\bar{\Lambda}_c\gamma_5ND+H.c.,\label{eq1}
\end{align}
where the coupling constant $g_{\Lambda_cND}=-13.98$ is established from the SU(4) invariant Lagrangians~\cite{Dong:2010xv} in terms of $g_{\pi{}NN}=13.45$ and
$g_{\rho{}NN}=6.0$.  $N$, $D$, and $\Lambda_c$ are the nucleon, $D$ meson, and $\Lambda_c^{+}$ baryon fields, respectively.

To calculate the diagrams depicted in Fig.~(\ref{cc1}), it is also to determine the effective Lagrangian densities for the interaction vertex involving $\bar{T}^{a}_{c\bar{s}0}(2900)^{--}K^{-}D^{-}$.  Since the spin-parity of $T^a_{c\bar{s}0}$ is established as $I(J^P)=1(0^{+})$, the coupling between
$T^a_{c\bar{s}0}$ and $KD$ predominantly occurs through $S$- and $D$-wave interactions. Given our focus on studying the production rate of
$\bar{T}^{a}_{c\bar{s}0}(2900)^{--}$ near the threshold region, the contribution from the lowest angular momentum state is most significant. This phenomenon
can be attributed to the higher energy requirement for $D$-wave production cross-section than that of the $S$-wave production cross-section.  Thus, in this
study, we will employ the effective Lagrangian densities corresponding to the $S$-wave $\bar{T}^{a}_{c\bar{s}0}(2900)^{--}K^{-}D^{-}$ interaction vertex.
It is important to highlight that $S$-wave effective Lagrangians are always characterized by fewer derivatives.  This leads us to express the Lagrangian
densities for the $S$-wave coupling between $\bar{T}^{a}_{c\bar{s}0}(2900)^{--}$ and $K^{-}D^{-}$ as follows~\cite{Zhu:2019vnr}
\begin{align}
{\cal{L}}_{\bar{T}_{c\bar{s}0}} &= g_{\bar{T}_{c\bar{s}0}}\bar{K}\vec{\tau}\cdot\bar{T}_{c\bar{s}0}\bar{D}\label{eq66},
\end{align}
where the $\tau$ is the corresponding pauli matrix reflecting the isospin of the $\bar{T}^{a}_{c\bar{s}0}(2900)$.  Please note that we have
\begin{align}
\vec{\tau}\cdot\vec{T}^{a}_{c\bar{s}0}=\left(
\begin{array}{cc}
T^{a}_{c\bar{s}0}(2900)^{+}         & \sqrt{2}T^{a}_{c\bar{s}0}(2900)^{++} \\
\sqrt{2}T^{a}_{c\bar{s}0}(2900)^{0} & -T^{a}_{c\bar{s}0}(2900)^{+}
\end{array}
\right)\label{eq5}
\end{align}
where the state $T^{a}_{c\bar{s}0}(2900)^{+}$ has not yet been discovered.  It only indicates the signal of $T^{a}_{c\bar{s}0}(2900)^{+}$ in
the $B^0\to{}D^{-}D^0K^{+}$ reaction~\cite{Duan:2023qsg}.

In Eq.~(\ref{eq66}),  the coupling constant $g_{\bar{T}_{c\bar{s}0}}$ is determined from the partial decay width of $T^{++}_{c\bar{s}0}\to{}K^{+}D^{+}$,
which is obtained as follow,
\begin{align}
\Gamma_{T^{a}_{c\bar{s}0}(2900)^{++}\to{}K^{+}D^{+}}=\frac{g^2_{T_{c\bar{s}0}}}{4\pi}\frac{|\vec{p}^{c.m}_{D^{+}}|}{m^2_{T^{++}_{c\bar{s}0}}},
\end{align}
where $\vec{p}^{c.m}_{D^{+}}$ is the three-vector momentum of the $D^{+}$ in the $T^{a}_{c\bar{s}0}(2900)^{++}$ meson rest frame.  Unfortunately, there is no
experimental information on the decay widths for $\Gamma{(T^{a}_{c\bar{s}0}(2900)^{++}\to{}K^{+}D^{+})}$, as this is very difficult to determine.  Thus, it is
necessary to rely on theoretical predictions, such as those of Refs.~\cite{Liu:2022hbk,Lian:2023cgs}.  Assuming  the $T^{a}_{c\bar{s}0}(2900)^{++}$ as a compact
multi-quark state, the partial decay width of the $T^{a}_{c\bar{s}0}(2900)^{++}\to{}K^{+}D^{+}$ is predicted to be $\Gamma({T^{++}_{c\bar{s}0}\to{}K^{+}D^{+}})=56.8\pm{}33.4$ MeV~\cite{Lian:2023cgs}. Using the corresponding experimental masses of the relevant particles given in Ref.~\cite{ParticleDataGroup:2022pth}, we obtain $g_{T_{c\bar{s}0}}=2.836_{-1.016}^{+0.739}$.  Note that the partial decay width of the $T^{a}_{c\bar{s}0}(2900)^{++}\to{}K^{+}D^{+}$ is also evaluated in
Ref.~\cite{Liu:2022hbk}, adopting the compact multi-quark state assignment for $T^{a}_{c\bar{s}0}(2900)^{++}$, and found that the obtained partial decay width
falls within the range reported in Ref.~\cite{Lian:2023cgs}.

\begin{figure}[h!]
\begin{center}
\includegraphics[bb=60 400 950 710, clip, scale=0.43]{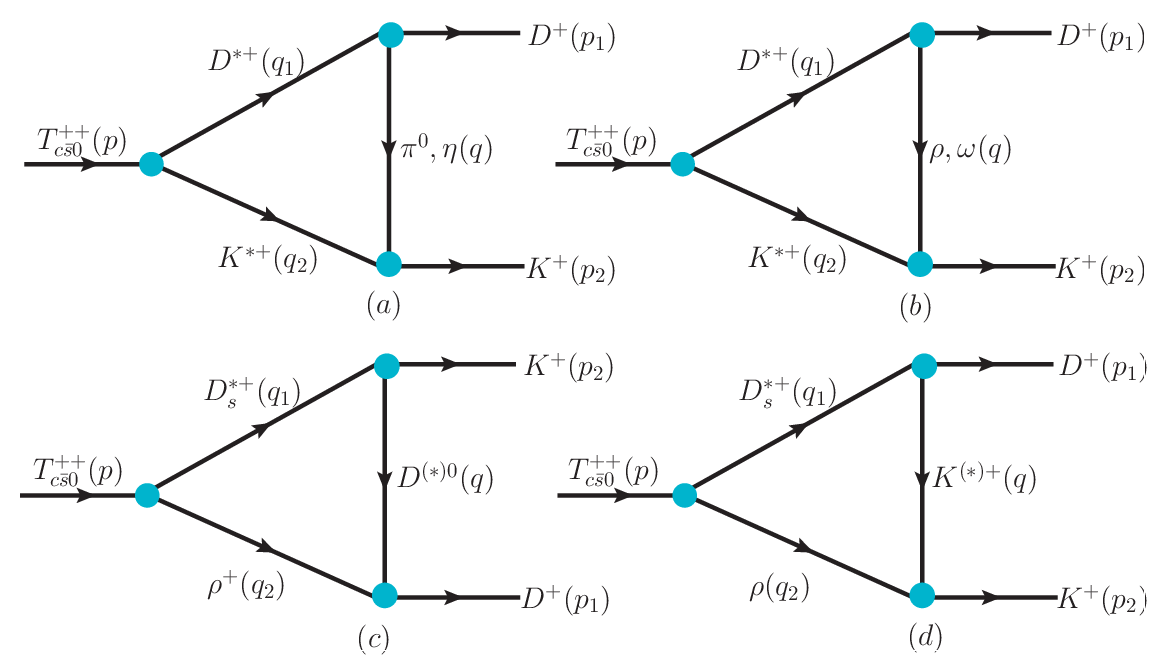}
\caption{Feynman diagram contributing to $T^{a}_{c\bar{s}0}(2900)^{++}\to{}K^{+}D^{+}$. We also show the definition of the kinematics ($q, p_1, p_2, q_1, q_2$)
that we use in the present calculation.}\label{cc3}
\end{center}
\end{figure}
Considering $T^{a}_{c\bar{s}0}(2900)^{++/0}$ as an $S$-wave $D^{}K^{}-D_s^{*}\rho$ molecule~\cite{Duan:2023lcj}, we
analyze the partial decay width of $T^{a}_{c\bar{s}0}(2900)^{++}$ into the $K^{+}D^{+}$ final state through hadronic loops with the help of the effective
Lagrangians.  The loop diagrams are depicted in Fig.~(\ref{cc3}), and the resulting partial decay widths are provided in Table.~\ref{tab12} (additional details
can be found in the Appendix.~\ref{secA1}).  Utilizing these decay widths, coupling constants are evaluated and collected in Table.~\ref{tab12}.
\begin{table}[h!]
\begin{center}
\caption{Coupling constants $g_{T_{\bar{c}s0}}$ and the $K^{+}D^{+}$ decay width (in units of MeV) of the $T^{a}_{c\bar{s}0}(2900)^{++/0}$.  The pole positions and effective couplings evaluated in Ref.~\cite{Duan:2023lcj}.}\label{tab12}
\begin{tabular}{cccccc} 		 	
  \hline\hline
  ~~~$\sqrt{s}_{pole}$     ~~~~~ &$|g_{D^{*}K^{*}}|$     ~~~~~& $|g_{D_s^{*}\rho}|$    ~~~~~& $\Gamma({T^{++}_{c\bar{s}0}\to{}K^{+}D^{+}})$  &~~~~~$g_{T_{\bar{c}s0}}$ \\\hline		
  ~~~2885                  ~~~~~ &5531                   ~~~~~& 5379                    ~~~~~& 48.72-54.59                                   &2.647-2.802    \\
  ~~~2887                  ~~~~~ &2198                   ~~~~~& 2082                    ~~~~~& 7.37-8.25                                     &1.029-1.089  \\
  \hline\hline
  \end{tabular}
 \end{center}
\end{table}

Due to the hadrons are not pointlike particles, it becomes imperative to incorporate form factors when evaluating the scattering amplitudes of the $K^{-}n\to{}\bar{T}^{a}_{c\bar{s}0}(2900)^{--}\Lambda^{+}_c$ reaction.  For the $t$-channel $D^{-}$ meson exchange diagram, we adopt a widely used approach
found in many previous studies~\cite{Zhu:2019vnr,He:2011jp,Xie:2015zga} with the expression
\begin{align}
{\cal{F}}_{D^{-}}(q^2_{D^{-}},m_{D^{-}})=\frac{\Lambda_{D^{-}}^2-m_{D^{-}}^2}{\Lambda_{D^{-}}^2-q^2_{D^{-}}}\label{eq7},
\end{align}
where $q^2_{D^{-}}$ and $m_{D^{-}}$ represent the four-momentum and mass of the exchanged $D^{-}$ meson, respectively. The parameter $\Lambda_{D^{-}}$ serves
as a hard cutoff, directly linked to the size of the hadron. Empirically, $\Lambda_{D^{-}}$ should exceed the $m_{D^{-}}$ mass by several hundred MeV at least.
Therefore, we choose $\Lambda_{D^{-}}=m_{D^{-}}+\alpha\Lambda_{QCD}$, following the precedent set by prior works~\cite{Zhu:2019vnr,He:2011jp,Xie:2015zga}.
The parameter $\alpha$ reflects the nonperturbative property of QCD at the low-energy scale, which will be taken as a parameter and discussed later.

In this study, we will examine the impact of the $K^{-}n$ initial state interaction on the cross-section of $\bar{T}^{a}_{c\bar{s}0}(2900)^{--}$ production in the $K^{-}n\to{}\bar{T}^{a}_{c\bar{s}0}(2900)^{--}\Lambda^{+}_c$ reaction.  A straightforward approach utilized in Ref.~\cite{Lebiedowicz:2011tp} yields a satisfactory
representation of the existing experimental data for $K^{-}p$ and $K^{-}n$ scattering at high energies.  Consequently, we employ this methodology to estimate
the initial state interaction (ISI) in the $K^{-}n\to K^{-}n$ reaction at high energies.

\begin{figure}[h!]
\begin{center}
\includegraphics[bb=90 610 950 710, clip, scale=0.6]{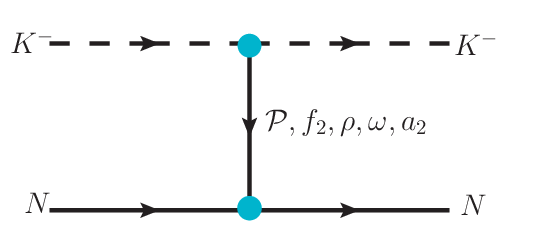}
\caption{Feynman diagram for the mechanism of the initial state interaction of the $K^{-}n\to{}K^{-}n$.}\label{cc2}
\end{center}
\end{figure}
The pertinent Feynman diagram depicting the ISI for $K^{-}N\to{}K^{-}N$ reaction is illustrated in Fig.~(\ref{cc2}), where exchanges involving the Pomeron and
$f_2$, $a_2$, $\rho$, and $\omega$ Reggeons (IR) are considered. The total amplitude ${\cal{T}}_{K^{-}N\to{}K^{-}N}$, incorporating Pomeron and Reggeon exchanges,
can be expressed as a sum of individual contributions, as indicated by \cite{Lebiedowicz:2011tp}
\begin{align}
{\cal{T}}_{K^{-}N\to{}K^{-}N}(s,t)&={\cal{A}}_{IP}(s,t)+{\cal{A}}_{f_2}(s,t)\pm{\cal{A}}_{a_2}(s,t)\nonumber\\
                                  &+{\cal{A}}_{\omega}(s,t)\pm{\cal{A}}_{\rho}(s,t),
\end{align}
where $s=(p_1+p_2)^2$ and $t=(p_1-q_1)^2$.  The $(+)$ and $(-)$ are for the $K^{-}p\to{}K^{-}p$ and $K^{-}n\to{}K^{-}n$ interactions, respectively.  For large center-of-mass
energies $\sqrt{s}$, the individual contribution to the $\bar{K}N\to{}\bar{K}N$ amplitude can be parameterized as follows
\begin{align}
{\cal{A}}_{i}(s,t)=\eta_is{\cal{C}}_i^{KN}(\frac{s}{s_0})^{\alpha_i(t)-1}\exp(\frac{{\cal{B}}^{i}_{KN}}{2}t),
\end{align}
where $i=IP$ represents the Pomeron and $f_2$, $a_2$, $\omega$, and $\rho$ Reggeons.  The energy scale is $s_0=1$ GeV$^2$. The coupling constants ${\cal{C}}_i^{\bar{K}N}$,
the parameters of the Regge linear trajectories $\alpha_i(t)=\alpha_i(0)+\alpha_i^{'}t$, the signature factors $\eta_i$, and the ${\cal{B}}^{i}_{\bar{K}N}$ utilized in Ref.\cite{Lebiedowicz:2011tp} offer a suitable description of the experimental data.  The parameters determined in Ref.\cite{Lebiedowicz:2011tp} are outlined in
Table~\ref{tab11}.
\begin{table}[h!]
\begin{center}
\caption{Parameters of Pomeron and Reggeon exchanges determined from elastic and total cross sections in Ref.~\cite{Lebiedowicz:2011tp}.}\label{tab11}
\begin{tabular}{cccccc} 		 	
  \hline\hline
  ~\textbf{i}~~ &$\eta_i$   ~~& $\alpha_{i}(t)$            ~~& ${\cal{C}}_{i}^{KN}$(mb)  ~~& ${\cal{B}}_i^{KN}$(GeV$^{-2}$)~ \\\hline		
  ~$IP$      ~~ &$i$        ~~& 1.081+(0.25~GeV$^{-2}$)t   ~~& $11.82$                   ~~&                   $5.5$       ~ \\
  ~$f_2$     ~~ &$-0.861+i$ ~~& 0.548+(0.93~GeV$^{-2}$)t   ~~& $15.67$                   ~~&                   $4.0$       ~ \\
  ~$\rho$    ~~ &$-1.162-i$ ~~& 0.548+(0.93~GeV$^{-2}$)t   ~~& $2.05$                    ~~&                   $4.0$       ~ \\
  ~$\omega$  ~~ &$-1.162-i$ ~~& 0.548+(0.93~GeV$^{-2}$)t   ~~& $7.055$                   ~~&                   $4.0$       ~ \\
  ~$a_2$     ~~ &$-0.861+i$ ~~& 0.548+(0.93~GeV$^{-2}$)t   ~~& $1.585$                   ~~&                   $4.0$       ~ \\
  \hline\hline
  \end{tabular}
 \end{center}
\end{table}

Using the effective Lagrangians mentioned above and taking the ISI of the $K^{-}p$ system into account, the full amplitude of the $K^{-}n\to{}\bar{T}^{a}_{c\bar{s}0}(2900)^{--}\Lambda^{+}_c$ reaction can be derived as
\begin{align}
{\cal{M}}^{full}={\cal{M}}^{Born}+{\cal{M}}^{K^{-}n-ISI},
\end{align}
where the Born amplitude is written as
\begin{align}
{\cal{M}}^{Born}&=if_Ig_{\bar{T}_{c\bar{s}0}}g_{\Lambda_cND}\bar{u}(q_2,s_{\Lambda^{+}_c})\gamma_5u(p_2,s_p)\nonumber\\
                &\times\frac{i}{(q_1-p_1)^2-m^2_{D^{-}}}{\cal{F}}_{D^{-}}[(q_1-p_1)^2_{D^{-}},m_{D^{-}}],
\end{align}
and the corrections to the Born amplitude due to $K^{-}n$ interactions were taken into account in~\cite{Lebiedowicz:2011tp,Lebiedowicz:2011nb}
as
\begin{align}
{\cal{M}}^{K^{-}n-ISI}=\frac{i}{16\pi^2s}\int{}d^2\vec{k}_t{\cal{T}}_{K^{-}n\to{}K^{-}n}(s,k^2_t){\cal{M}}^{Born}.
\end{align}
where $k_t$ is the momentum transfer in the $K^{-}n\to{}K^{-}n$ reaction.  $\bar{u}(q_2,s_{\Lambda^{+}_c})$ and $u(p_2,s_p)$ are the Dirac spinors,
with $s_{\Lambda^{+}_c}(q_2)$ and $s_p(p2)$ being the spins (the four-momenta) of the outgoing $\Lambda^{+}_c$ and the initial proton, respectively.

With the scattering amplitudes of the
$K^{-}n\to{}\bar{T}^{a}_{c\bar{s}0}(2900)^{--}\Lambda^{+}_c$ reaction obtained in the previous section, the differential cross section in the center of mass
(cm) frame for the process $K^{-}n\to{}\bar{T}^{a}_{c\bar{s}0}(2900)^{--}\Lambda^{+}_c$ can be calculated~\cite{ParticleDataGroup:2022pth}
\begin{align}
\frac{d\sigma}{d\cos\theta}=\frac{m_Nm_{\Lambda_c^{+}}}{8\pi{}s}\frac{|\vec{q}_{1cm}|}{|\vec{p}_{1cm}|}(\frac{1}{2}\sum_{s_{n},s_{\Lambda_c^{+}}}|{\cal{M}}|^2),
\end{align}
where the $\theta$ is the scattering angle of the outgoing  $\bar{T}^{a}_{c\bar{s}0}$ meson relative to the beam direction, while $\vec{p}^{1cm}$ and $\vec{q}^{1cm}$
are the $K^{-}$ and $\bar{T}^{a}_{c\bar{s}0}$ meson three momenta in the center-of-mass frame, respectively, which are
\begin{align}
|\vec{p}_{1cm}|=\frac{\lambda^{1/2}(s,m_{K^{-}},m_{n})}{2\sqrt{s}};~
|\vec{q}_{1cm}|=\frac{\lambda^{1/2}(s,m_{T^{a}_{c\bar{s}0}},m_{\Lambda_c^{+}})}{2\sqrt{s}}.
\end{align}
Here $\lambda(x,y,z)=(x-y-z)^2-4yz$ is the K\"{a}llen function.

\section{RESULTS AND DISCUSSIONS}\label{Sec: results}
With the formalism and ingredients given above, the cross-section as a function of the beam momentum $P_{K^{-}}$ for $K^{-}n\to{}\bar{T}^{a}_{c\bar{s}0}
(2900)^{--}\Lambda^{+}_c$  reaction can be easily obtained.  Before presenting the results, it is important to discuss the parameter $\alpha$ that relates
to the form factor. This is because the value of the cross-section is highly sensitive to the model parameter $\alpha$.  However, determining the value of
$\alpha$ from first principles is currently not feasible.
Instead, it can be better determined from the experimental data.  Indeed, it has been
established that the free parameter $\alpha=1.5$ or $1.7$ were fixed by fitting the experimental data of the processes $e^{+}e^{-}\to{}D\bar{D}$~\cite{Belle:2007qxm}
and $e^{+}e^{-}\to\gamma{ISR}D\bar{D}$~\cite{BaBar:2006qlj}. The procedures for this fitting is outlined in Ref.\cite{Guo:2016iej}.  For this study,
we adopt the values $\alpha=1.5$ or $1.7$, as they have been determined from the experimental data of Refs.~\cite{Belle:2007qxm,BaBar:2006qlj} using
the same $D$ form factors employed in our current work.
\begin{figure}[h!]
\begin{center}
\includegraphics[bb=10.0 140 950 480, clip, scale=0.55]{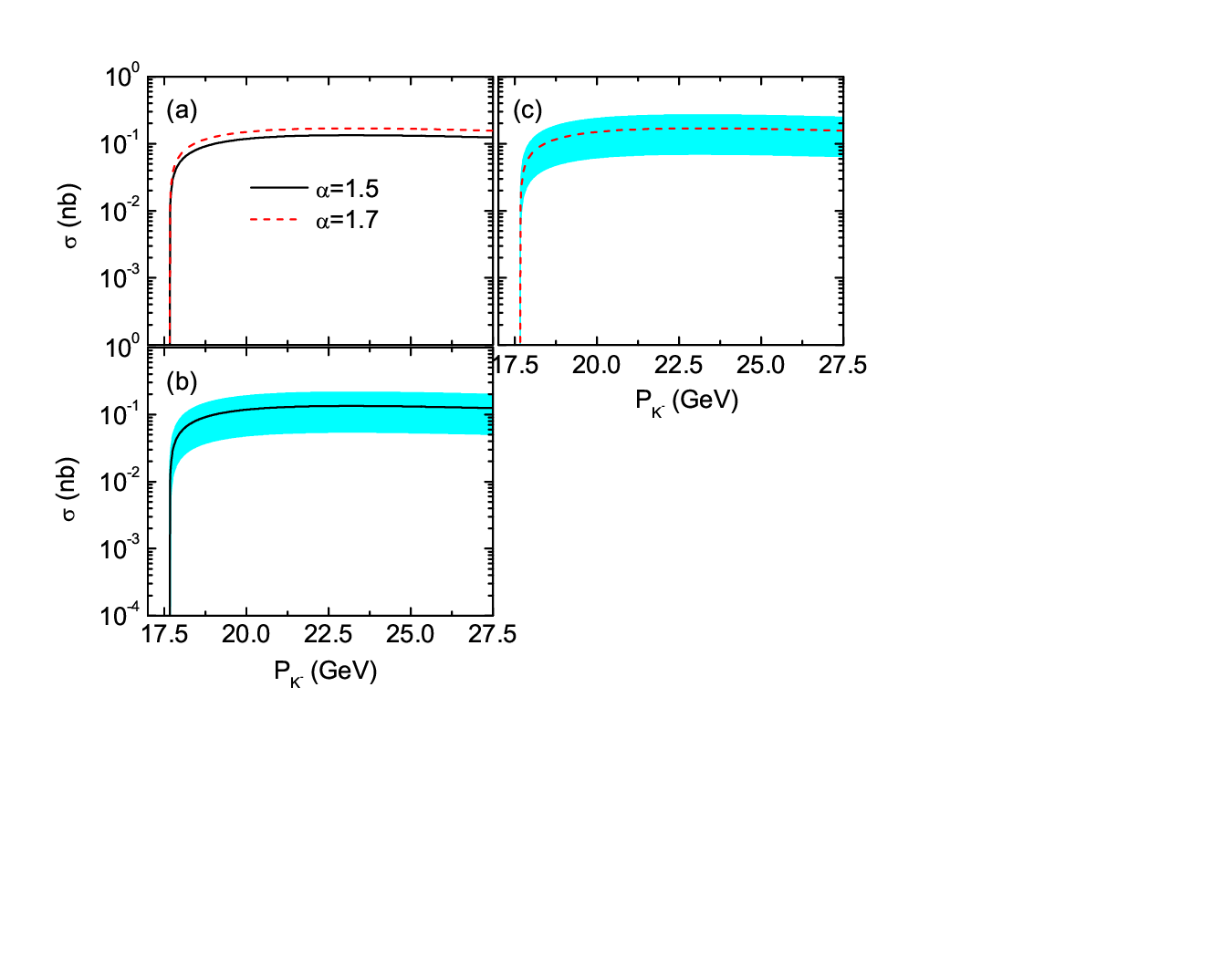}
\caption{The cross-section for the $K^{-}n\to{}\bar{T}^{a}_{c\bar{s}0}(2900)^{--}\Lambda^{+}_c$ reaction as a function of the beam momentum $P_{K^{-}}$
with assuming $\bar{T}^{a}_{c\bar{s}0}$ as a compact tetraquarks  state.  The (a) corresponding to the coupling constant $g_{T_{c\bar{s}0}}=2.836$, and
the cyan bands denote the cross-sections for different $\Gamma(T^{++}_{c\bar{s}0}\to{}K^{+}D^{+})$ values~\cite{Lian:2023cgs}.}\label{cc12}
\end{center}
\end{figure}

With the obtained $\alpha$ value, the cross-section for $K^{-}n\to{}\bar{T}^{a}_{c\bar{s}0}(2900)^{--}\Lambda^{+}_c$ reaction is evaluated by treating
$T_{c\bar{s}0}^{a}$ as a compact multi-quark state.  The theoretical results obtained with a cutoff $\alpha=1.5$ or $1.7$ for the beam energy from near
threshold up to 27.5 GeV are shown in Fig.~\ref{cc12}.  We can find that the cross-section exhibits a sharp increase near the threshold of $\bar{T}^{a}_{c\bar{s}0}(2900)^{--}\Lambda^+_c$, an effect attributed to the opening of phase space at that energy.  Following this, the cross-section
continues to increase, albeit at a comparatively slower rate compared to the threshold region. However, a modest decline in the cross-section is observed
as the beam energy $P_{K^-}$ is varied from 23.1 to 27.5 GeV.  For deeper insight, we illustrate the obtained total cross-section behavior, ranging from
approximately 0.134 nb to 0.127 nb for $\alpha=1.5$, with varying beam momentum from 23.25 GeV to 26.80 GeV. Within the same energy range, but adopting
$\alpha=1.7$, the cross-section spans from 0.169 nb to 0.160 nb.  These outcomes suggest that the value of the cross-section is not very sensitive to the
model parameter $\alpha$ when varying the cutoff parameter $\alpha$ from 1.5 to 1.7.

In addition to showing the central values of the cross-sections corresponding to $g_{T_{c\bar{s}0}}=2.836$, we also present the variation of the cross-sections for
different $g_{T_{c\bar{s}0}}$ values, which are determined based on the theoretically predicted partial decay width of $T^{++}_{c\bar{s}0}\to{}K^{+}D^{+}$~\cite{Lian:2023cgs}.
We depict the results for the cutoffs $\alpha=1.5$ and $\alpha=1.7$ in the Fig.\ref{cc12} (b) and Fig.~\ref{cc12} (c), respectively.  Remarkably, a significant
variations in the cross-sections is observed. For $\alpha=1.5$, the obtained cross-section ranges from 0.052 nb to 0.201 nb, and for $\alpha=1.7$, it ranges from
0.066 nb to 0.253 nb, both at an example energy of approximately $P_{K^{-}}=26.80$ GeV.  This suggests that the cross-section for the maximum value is about four
times larger than that of the minimum value.  This discrepancy can be attributed to the fact that the ratio of the coupling constant for the maximum value to that
of the minimum value is around 2.0. And the cross-section is proportional to the square of the coupling constant.

\begin{figure}[h!]
\begin{center}
\includegraphics[bb=20 80 950 355, clip, scale=0.45]{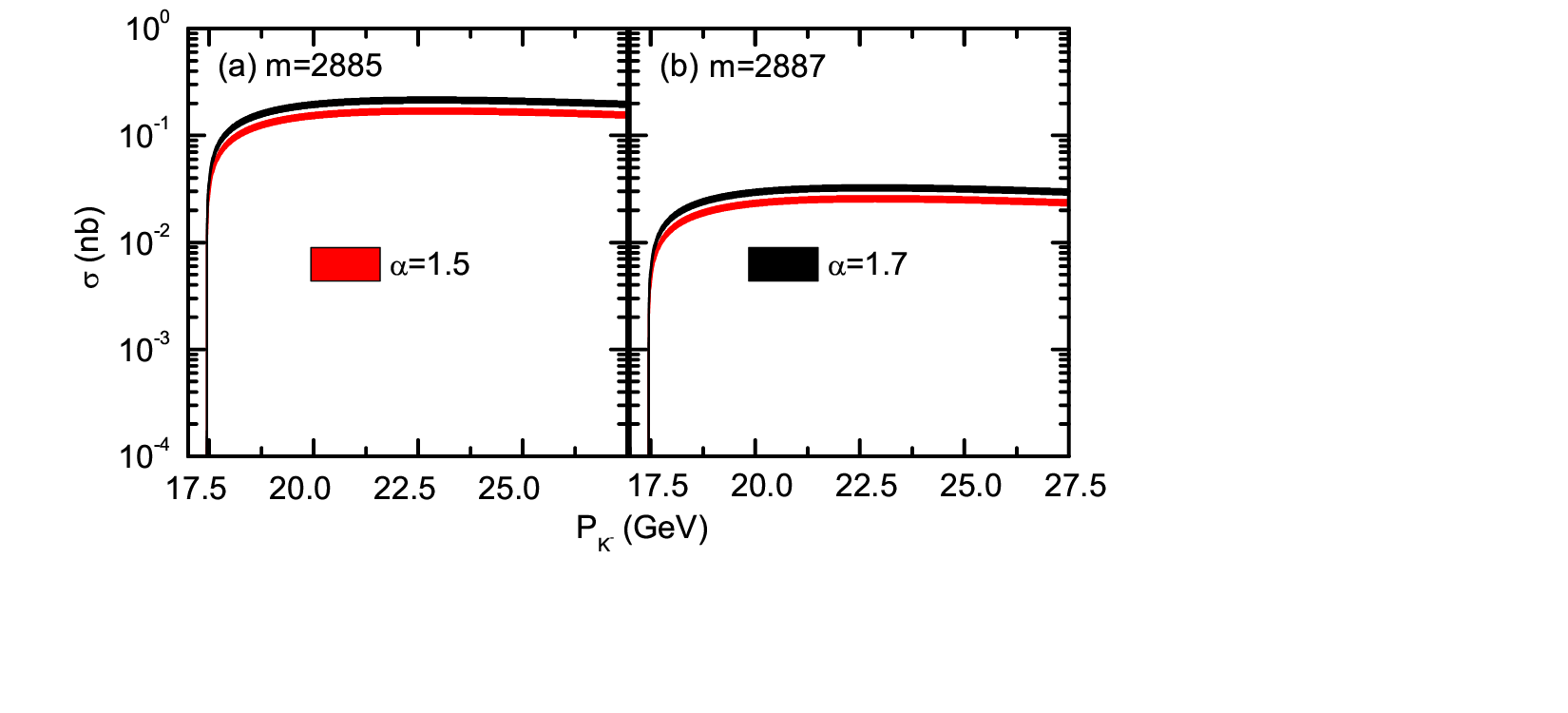}
\caption{The cross-section for the $K^{-}n\to{}\bar{T}^{a}_{c\bar{s}0}(2900)^{--}\Lambda^{+}_c$ reaction as a function of the beam momentum $P_{K^{-}}$
with assuming $\bar{T}^{a}_{c\bar{s}0}$ as a $K^{*}D^{*}-D_s^{*}\rho$  molecule. The color bands denote the cross-sections for different $\Gamma(T^{++}_{c\bar{s}0}\to{}K^{+}D^{+})$ values, and $m$ is the masses of the bound states $T^{a}_{c\bar{s}0}(2900)$ that obtained in Ref.~\cite{Duan:2023lcj}. }\label{cc13}
\end{center}
\end{figure}

We now shift our focus to the cross-section of the $K^{-}n\to{}\bar{T}^{a}_{c\bar{s}0}(2900)^{--}\Lambda^{+}_c$ reaction, considering $T^{a}_{c\bar{s}0}(2900)$ as a $K^{*}D^{*}-D_s^{*}\rho$ molecule. The cross-section, varying with the beam energy $P_{K^{-}}$ from just above the threshold up to 27.5 GeV, is depicted in Fig.~\ref{cc13}.
We clearly observe that the line shapes of the cross-sections mirror those obtained by considering $T^{a}_{c\bar{s}0}(2900)$ as a compact multi-quark state. That means
the cross-section also increases sharply near the threshold, followed by a gradual and sustained increase at higher energies, concluding with a gradual decrease.

The results in Fig.~\ref{cc13} also tell us that the $\bar{T}^{a}_{c\bar{s}0}(2900)^{--}$ production cross-section for the cutoff $\alpha=1.5$ is slightly smaller than
that for cutoff $\alpha=1.7$. The variation of the cross-sections for different $\Gamma(T^{++}_{c\bar{s}0}\to{}K^{+}D^{+})$ values is very small. To see how much the
cross-section depends on the $\Gamma(T^{++}_{c\bar{s}0}\to{}K^{+}D^{+})$ decay widths and the cutoff $\alpha$, we take the cross-section at a beam momentum of about
$P_{K^{-}}=25.0$ GeV and the mass of the bound state $m=2885$ MeV as an example.  The so-obtained cross-section ranges from 0.156 nb to 0.175 nb for $\alpha=1.5$
and from $0.197$ to 0.221 nb for $\alpha=1.7$.  We also find that if the $\bar{T}^{a}_{c\bar{s}0}(2900)$ is produced as a $K^{*}D^{*}-D_s^{*}\rho$ molecule with mass
$m=2885$ MeV, the cross-section is significantly larger than the results obtained by assuming $\bar{T}^{a}_{c\bar{s}0}(2900)$ as a $K^{}D^{}-D_s^{*}\rho$ molecule with
mass $m=2887$ MeV, by about $6-8$ times.  In other words, the cross-section is heavily depends on the masses of the bound states.

By comparing the cross-sections depicted in Fig.\ref{cc12} and Fig.\ref{cc13}, we observe that if the $\bar{T}^{a}_{c\bar{s}0}(2900)^{--}$ is a compact tetraquark state,
its production cross-sections match the results predicted by considering the $\bar{T}^{a}_{c\bar{s}0}(2900)^{--}$ as a $K^{*}D^{*}-D_s^{*}\rho$ molecule with a mass of $m=2885$
MeV.  Specifically, the cross-sections for these two assignments can reach 0.269 nb (for compact tetraquark state) and 0.227 nb (for $K^{*}D^{*}-D_s^{*}\rho$ molecule),
respectively.  These results suggest that if the $\bar{T}^{a}_{c\bar{s}0}(2900)$ is a $K^{*}D^{*}-D_s^{*}\rho$ molecule with a mass of $m=2887$ MeV,  distinguishing its $K^{*}D^{*}-D_s^{*}\rho$ molecular nature from the compact tetraquark state is challenging through the $K^{-}n\to{}\bar{T}^{a}_{c\bar{s}0}(2900)^{--}\Lambda^{+}_c$
reaction.  However, when considering a smaller cross-section obtained from the assumption that the $\bar{T}^{a}_{c\bar{s}0}(2900)^{--}$ is a $K^{*}D^{*}-D_s^{*}\rho$ molecule
with a mass of $m=2887$ MeV, the cross-section is limited to 0.034 nb. This discrepancy magnifies the difference between the results of the two scenarios by a factor of
about 8.0.   This indicates that if the mass of the $K^{*}D^{*}-D_s^{*}\rho$ molecule is $m=2887$ MeV, a clear conclusion about the the nature of the
$\bar{T}^{a}_{c\bar{s}0}(2900)$ can be easily obtained by comparing the obtained cross-section with future experimental data.

\begin{figure}[h!]
\begin{center}
\includegraphics[bb=-50 90 950 410, clip, scale=0.45]{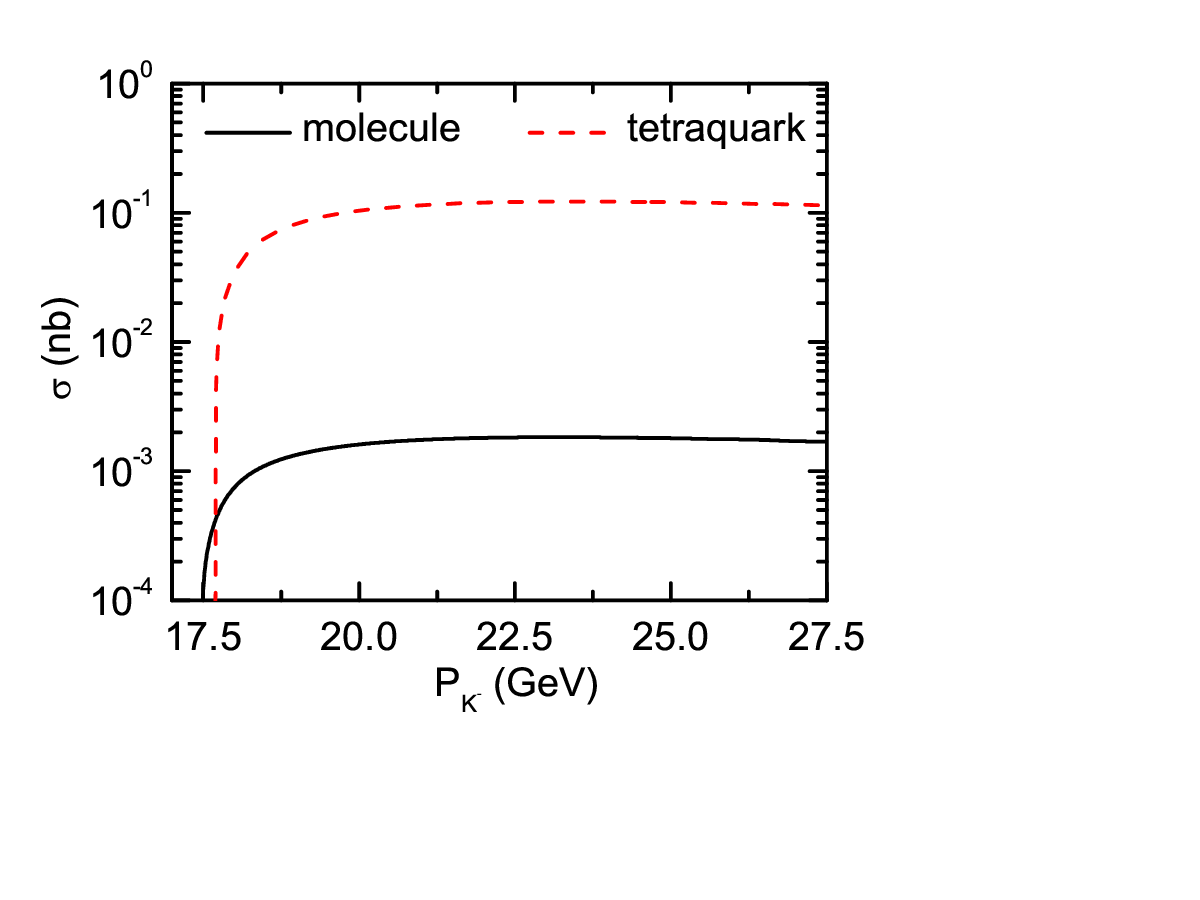}
\caption{The cross-sections for the $K^{-}n\to{}\bar{T}^{a}_{c\bar{s}0}(2900)^{--}\Lambda^{+}_c\to{}\pi^{-}D^{-}_s\Lambda^{+}_c$
reaction as a function of the beam momentum $P_{K^{-}}$ with assuming $\bar{T}^{a}_{c\bar{s}0}$ as a $K^{*}D^{*}-D_s^{*}\rho$ molecule (Black solid line)
or compact tetraquark state (red dash line). Here, we display results associated with $\alpha=1.7$, the bound state mass $m=2885$ MeV, and the maximum value
of the coupling constants $g_{T^{a}_{c\bar{s}0}}$.}\label{cc17}
\end{center}
\end{figure}
However, a more distinct comparison can be drawn from the production of $\bar{T}^{a}_{c\bar{s}0}(2900)$ in the $K^{-}n\to{}\bar{T}^{a}_{c\bar{s}0}(2900)^{--}\Lambda^{+}_c\to{}\pi^{-}D^{-}_s\Lambda^{+}_c$ reaction, and the results of this comparison are shown in Fig.~\ref{cc17}.
We can find that the cross-section for the molecule assignment of $\bar{T}^{a}_{c\bar{s}0}(2900)$ can reach up to $1.83\times{}10^{-3}$ nb.  This value is significantly
smaller than that of 0.122 nb at the same beam energy.  And the larger cross-section was derived from considering the $\bar{T}^{a}_{c\bar{s}0}(2900)$ as a compact tetraquark
state. The significant difference between the results in these two pictures arises from considering $T^{a}_{c\bar{s}0}(2900)$ in two different ways.  If it's seen as a
compact tetraquark state, the calculated partial decay width of $T^{a}_{c\bar{s}0}(2900)^{++}\to{}\pi^{+}D^{+}_s$ is $48.5\pm{}30.0$ MeV~\cite{Lian:2023cgs}(we use 78.5 MeV).
This contrasts sharply with the range of 0.132-1.167 MeV (we use 1.167 MeV) obtained by treating $T^{a}_{c\bar{s}0}(2900)$ as a $K^{*}D^{*}-D_s^{*}\rho$ molecule with a mass
of $m$=2885 MeV.  Importantly, the partial decay width of $T^{a}_{c\bar{s}0}(2900)^{++}\to{}\pi^{+}D^{+}_s$ from the $D_s^{*}\rho$ channel contribution is prohibited, and
the necessary amplitudes required for the $T^{a}_{c\bar{s}0}(2900)^{++}\to{}\pi^{+}D^{+}_s$ reaction in our work are available in Ref.~\cite{Yue:2022mnf}. Note that the
cross-section for the $K^{-}n\to{}\pi^{-}D^{-}_s\Lambda^{+}_c$ reaction are computed from the following differential cross-section~\cite{Kim:2017nxg}
\begin{align}
\frac{d\sigma_{K^{-}n\to{}\pi^{-}D^{-}_s\Lambda^{+}_c}}{d{\cal{M}}_{\pi^{-}D_s^{-}}}&\approx{}\frac{2m_{\bar{T}^{a}_{c\bar{s}0}}{\cal{M}}_{\pi^{-}D_s^{-}}}{\pi}\nonumber\\
                                                                                    &\times\frac{\sigma_{K^{-}n\to{}\bar{T}^{a}_{c\bar{s}0}(2900)^{--}\Lambda^{+}_c}\Gamma_{\bar{T}^{a}_{c\bar{s}0}(2900)^{--}\to{}\pi^{-}D^{-}_s}}{({\cal{M}}^2_{\pi^{-}D_s^{-}}-m^2_{\bar{T}^a_{c\bar{s}0}})^2+m^2_{\bar{T}^a_{c\bar{s}0}}\Gamma^2_{\bar{T}^a_{c\bar{s}0}}},
\end{align}
where $m_{\bar{T}^{a}_{c\bar{s}0}}$ and $\Gamma_{\bar{T}^a_{c\bar{s}0}}$ are the mass and width of the $\bar{T}^{a}_{c\bar{s}0}$, respectively.  ${\cal{M}}_{\pi^{-}D_s^{-}}$
is in the range of the $(m_{\pi^{-}}+m_{D_s^{-}})-(\sqrt{s}-m_{\Lambda_c^{+}})$.

\begin{figure}[h!]
\begin{center}
\includegraphics[bb=-80 120 950 570, clip, scale=0.30]{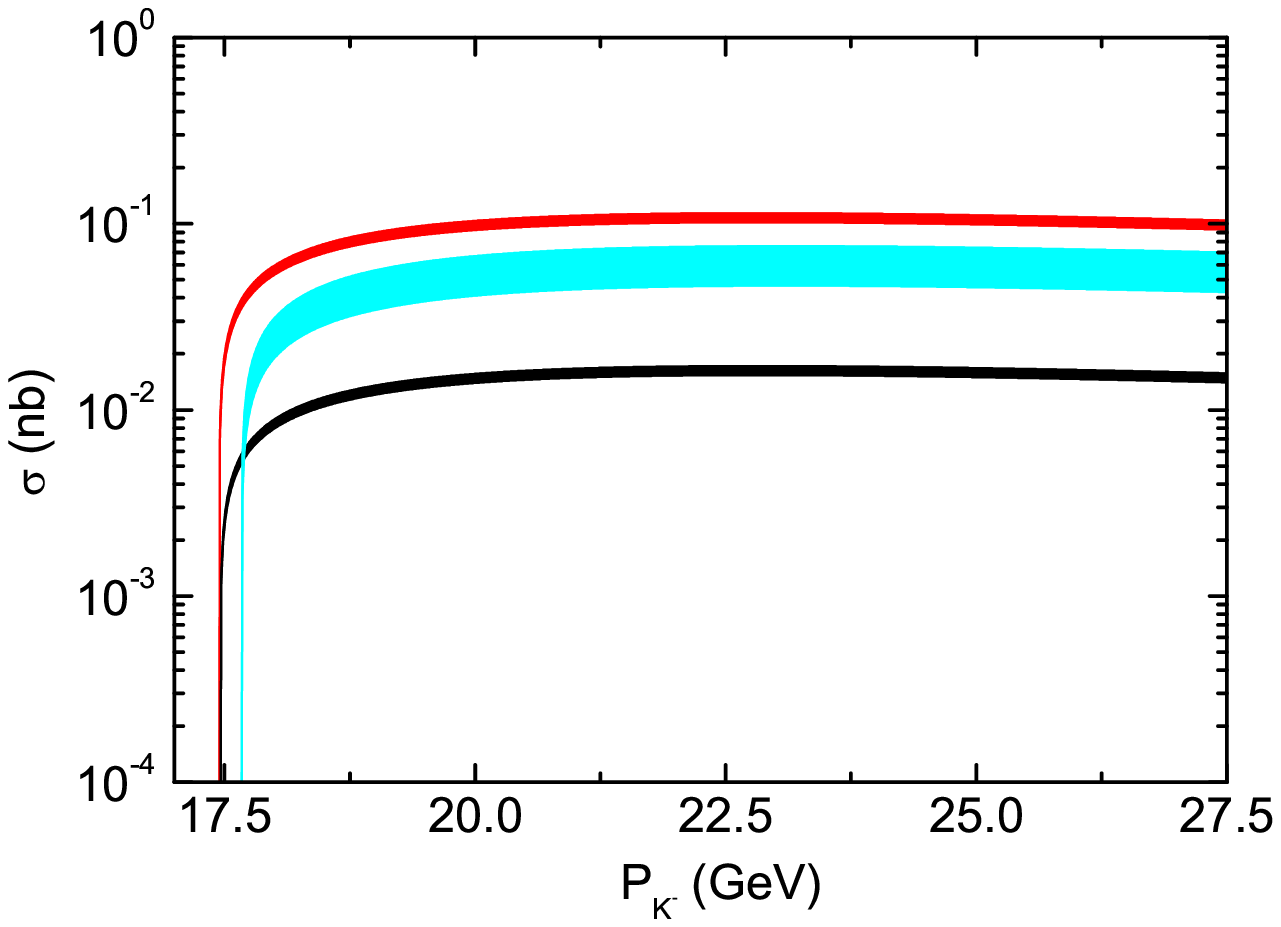}
\caption{The cross-section for the $K^{-}p\to{}\bar{T}^{a}_{c\bar{s}0}(2900)^{-}\Lambda^{+}_c$ reaction as a function of the beam momentum $P_{K^{-}}$
with assuming $\bar{T}^{a}_{c\bar{s}0}$ as a $K^{*}D^{*}-D_s^{*}\rho$ molecule (red and black bands) or compact tetraquark (cyan bands). The color bands
denote the cross-sections for different $\Gamma(T^{+}_{c\bar{s}0}\to{}K^{+}D^{0})$ values.  The red and black bands corresponding to the bound state mass
$m=2885$ MeV and $M$=2887 MeV, respectively, that obtained in Ref.~\cite{Duan:2023lcj}. The cyan bands is the cross-section for the mass of the bound state
$m=2921$ MeV~\cite{LHCb:2022sfr}.}\label{cc14}
\end{center}
\end{figure}
Considering isospin symmetry, the decay width of $K^{+}D^0$ for the unreported $T^{a}_{c\bar{s}0}(2900)^{+}$ state is expected to be half of the partial
decay width of $T^{++}_{c\bar{s}0}\to{}K^{+}D^{+}$ (as seen in Eq.~\ref{eq5}). If we regard the $T^{a}_{c\bar{s}0}(2900)^{+}$ as a $K^{*}D^{*}-D_s^{*}\rho$
molecule, the predicted partial decay widths for the bound state with masses $m=2885$ MeV and $m=2887$ MeV are $\Gamma({T^{+}_{c\bar{s}0}\to{}K^{+}D^{0}})=24.36-27.30$
MeV and $\Gamma({T^{+}_{c\bar{s}0}\to{}K^{+}D^{0}})=2.69-4.13$ MeV, respectively. Furthermore, the decay width $\Gamma({T^{+}_{c\bar{s}0}\to{}K^{+}D^{0}})$
is estimated to be within the range of 11.7-45.1 MeV when considering the $T^{+}_{c\bar{s}0}$ as a compact tetraquark state.  These predictions open up an
opportunity to search for the $\bar{T}^{a}_{c\bar{s}0}(2900)^{-}$ [$\bar{T}^{a}_{c\bar{s}0}(2900)^{-}$ is the antiparticle of $T^{a}_{c\bar{s}0}(2900)^{+}$]
in the $K^{-}p\to{}\bar{T}^{a}_{c\bar{s}0}(2900)^{-}\Lambda_c^{+}$ reaction.  The cross-section for the $K^{-}p\to{}\bar{T}^{a}_{c\bar{s}0}(2900)^{-}\Lambda_c^{+}$
reaction, with a cutoff $\alpha=1.7$ and beam momentum $P_{K^{-}}$ ranging from near threshold to 27.5 GeV, are presented in Fig.~\ref{cc14}.  The findings
reveal that the maximum value of the cross-section for $\bar{T}^{a}_{c\bar{s}0}(2900)^{-}$ production in the $K^{-}p\to{}\bar{T}^{a}_{c\bar{s}0}(2900)^{-}\Lambda_c^{+}$
reaction is approximately 0.114 nb, which is bigger than 0.075 nb that is the maximum values by assigning the $\bar{T}^{a}_{c\bar{s}0}(2900)^{-}$ as a compact
tetraquark state.

\begin{figure}[h!]
\begin{center}
\includegraphics[bb=50 80 950 430, clip, scale=0.40]{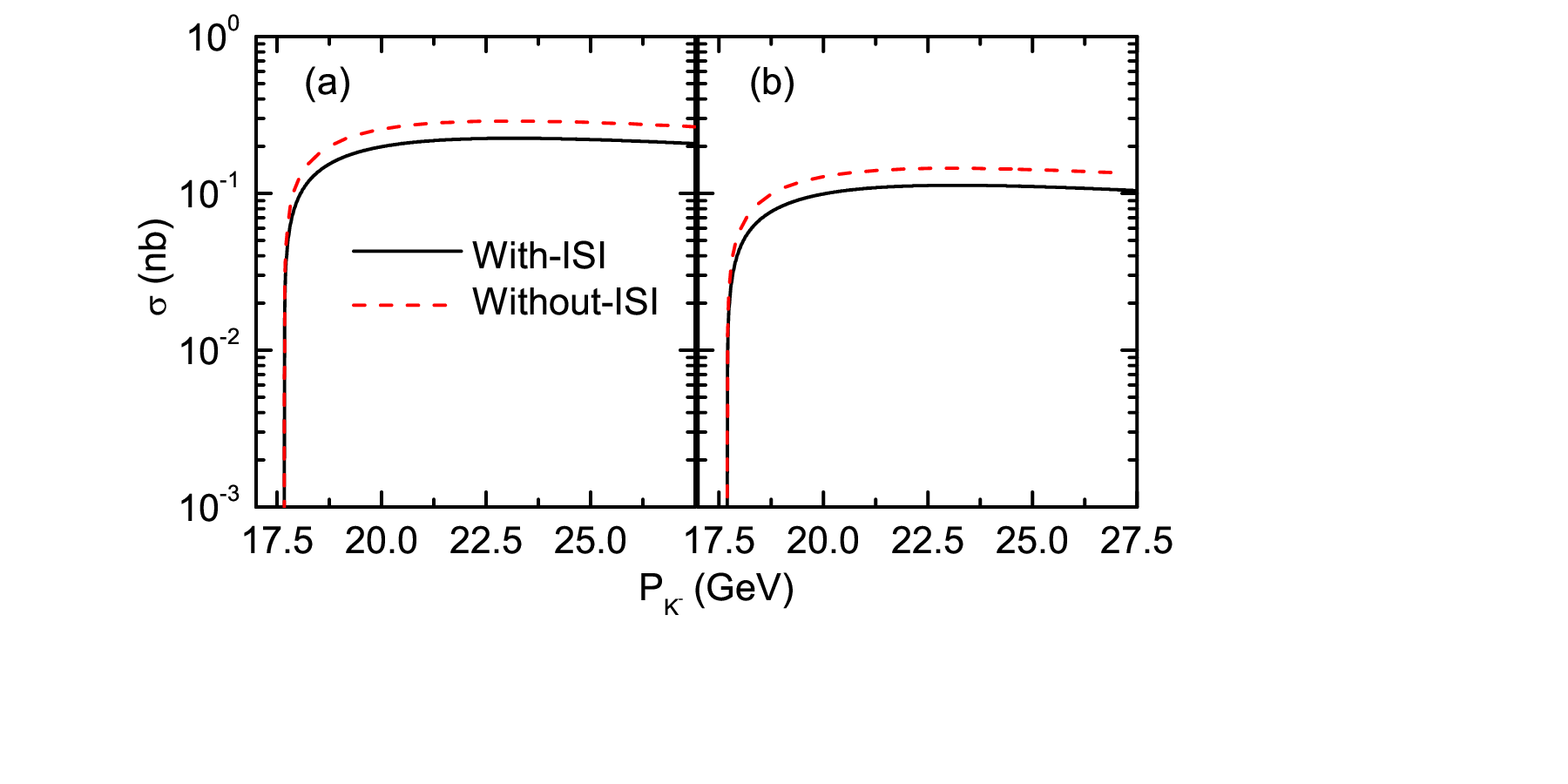}
\caption{The cross-sections for the $K^{-}p\to{}\bar{T}^{a}_{c\bar{s}0}(2900)^{-}\Lambda^{+}_c$ and $K^{-}p\to{}\bar{T}^{a}_{c\bar{s}0}(2900)^{-}\Lambda_c^{+}$
reactions as a function of the beam momentum $P_{K^{-}}$ with assuming $\bar{T}^{a}_{c\bar{s}0}$ as a compact tetraquark state.  The black solid lines and dash red
lines represent the cross-section that obtained with and without ISI, respectively. }\label{cc15}
\end{center}
\end{figure}
It is worth noting that the contributions to the $K^{-}p\to{}\bar{T}^{a}_{c\bar{s}0}(2900)^{-}\Lambda_c^{+}$ reaction is mediated by the exchange of $D$ meson in the
$t$-channel, which is identical to the production process of its isospin partner $\bar{T}^{a}_{c\bar{s}0}(2900)^{--}$.  The only difference is the influence of $\bar{K}N$ ISI
on their product cross-sections.  To show the effect from the $\bar{K}N$ ISI, we compare the cross-sections obtained with and without ISI for the cutoff $\alpha=1.7$ and
$g_{T_{c\bar{s}0}}=2.836$ in  Fig.~\ref{cc15}, for the $K^{-}n\to{}\bar{T}^{a}_{c\bar{s}0}(2900)^{--}\Lambda_c^{+}$ [Fig.~\ref{cc15}(a)] and  $K^{-}p\to{}\bar{T}^{a}_{c\bar{s}0}(2900)^{-}\Lambda_c^{+}$ [Fig.~\ref{cc15}(b)] reactions, respectively.  In Fig.~\ref{cc15}, the black solid line are the pure Born amplitude contribution, while the dash red line are the full results. We can find that the presence of $\bar{K}N$ ISI leads to a reduction in the cross-section of the $K^{-}p\to{}\bar{T}^{a}_{c\bar{s}0}(2900)^{-}\Lambda_c^{+}$ and $K^{-}p\to{}\bar{T}^{a}_{c\bar{s}0}(2900)^{-}\Lambda_c^{+}$ reactions by approximately 20\%.  This suggest
that the $\bar{K}N$ ISI have a significant impact on the search for $\bar{T}^{a}_{c\bar{s}0}(2900)$ in the $K^{-}N\to{}\bar{T}^{a}_{c\bar{s}0}(2900)^{-/--}\Lambda_c^{+}$
reactions.  Similar conclusions regarding the reduction in cross-section could be drawn if one were to consider the $\bar{T}^{a}_{c\bar{s}0}(2900)$ as a $K^{}D^{}-D_s^{*}\rho$ molecule, and we here do not discuss it in details.

\section{Summary}\label{sec:summary}
Theoretical investigations on the production processes will be helpful to distinguish which inner structure of the$T^{a}_{c\bar{s}0}(2900)$
is possible.  This is because the different production mechanisms of the $T^{a}_{c\bar{s}0}(2900)$ rely on its structure assignments.  In this
work, we examine the recently  observed $\bar{T}^{a}_{c\bar{s}0}(2900)$ production in the $K^{-}n\to{}\bar{T}^{a}_{c\bar{s}0}(2900)^{--}\Lambda^{+}_c$
and $K^{-}n\to{}\bar{T}^{a}_{c\bar{s}0}(2900)^{--}\Lambda^{+}_c\to{}\pi^{-}D^{-}_s\Lambda^{+}_c$ reactions by considering $T^{a}_{c\bar{s}0}$ as
a molecular state and a compact multiquark state, respectively.  The $T^{a}_{c\bar{s}0}(2900)$ can be produced though the exchange of $D$- meson
in the $t$-channel.

Using the coupling constants of the $\bar{T}^{a}_{c\bar{s}0}(2900)$ to $KD$ channel obtained from molecule or compact tetraquark picture of the
$\bar{T}^{a}_{c\bar{s}0}(2900)$, we compute the cross-sections for the $K^{-}n\to{}\bar{T}^{a}_{c\bar{s}0}(2900)^{--}\Lambda^{+}_c$ and $K^{-}n\to{}\bar{T}^{a}_{c\bar{s}0}(2900)^{--}\Lambda^{+}_c\to{}\pi^{-}D^{-}_s\Lambda^{+}_c$ reactions, respectively.  The numerical results reveal
that whether $\bar{T}^{a}_{c\bar{s}0}(2900)$ is categorized as a molecule or a compact tetraquark state, the cross-sections for the $K^{-}n\to{}\bar{T}^{a}_{c\bar{s}0}(2900)^{--}\Lambda^{+}_c$ reaction exhibit similar magnitudes, spanning roughly from 0.06 nb to 0.3 nb.
Nevertheless, a more distinct comparison can be drawn by computing the cross-section of the $K^{-}n\to{}\bar{T}^{a}_{c\bar{s}0}(2900)^{--}\Lambda^{+}c\to{}\pi^{-}D^{-}_s\Lambda^{+}_c$ process. The results indicate that,
assuming the molecule assignment for $\bar{T}^{a}_{c\bar{s}0}(2900)$, the cross-section for the $K^{-}n\to{}\bar{T}^{a}_{c\bar{s}0}(2900)^{--}\Lambda^{+}c\to{}\pi^{-}D^{-}_s\Lambda^{+}_c$ reaction could peak at $1.83\times{}10^{-3}$ nb, notably smaller than
the 0.122 nb obtained by presuming $\bar{T}^{a}_{c\bar{s}0}(2900)$ to be a compact tetraquark state. These findings are poised for future experimental
measurement and can serve as tests to discern the nature of $\bar{T}^{a}_{c\bar{s}0}(2900)$.  Lastly, we suggest a quest for the unreported charged
tetraquark $T^{a}_{c\bar{s}0}(2900)^{+}$ in the $K^{-}p\to{}\bar{T}^{a}_{c\bar{s}0}(2900)^{-}\Lambda^{+}_c$ reaction due to its production cross-section
can reach up to 0.114 nb.

A rough estimation finds that the production cross-section of the $\bar{T}^{a}_{c\bar{s}0}(2900)$ through high-energy proton-proton collisions is approximately 17 fb
(see Fig.3 in Ref.~\cite{LHCb:2022sfr}). This value is about $10^{4}$ times smaller than our calculated results. This difference is why we propose to search for the $\bar{T}^{a}_{c\bar{s}0}(2900)$ in the reactions $K^{-}p\to{}\bar{T}^{a}_{c\bar{s}0}(2900)^{-}\Lambda^{+}_c$ and $K^{-}n\to{}\bar{T}^{a}_{c\bar{s}0}(2900)^{--}\Lambda^{+}_c\to{}\pi^{-}D^{-}_s\Lambda^{+}_c$.
Furthermore, kaon beams with momenta ranging from 50 to 280 GeV/c and an RMS below 5\% are available from the M2 beamline at AMBER\cite{Quintans:2022utc}.
AMBER is a new fixed-target experiment at CERN that began its data-taking in 2023, providing an excellent platform for the search for $\bar{T}^{a}_{c\bar{s}0}(2900)$.

\section*{Acknowledgments}
This work was supported by the National Natural Science Foundation
of China under Grant No.12005177.

\section*{Appendix: PARTIAL DECAY WIDTHS $T^a_{c\bar{s}0}(2900)^{++}\to{}K^{+}D^{+}$}\label{secA1}
In this Appendix, we show how to compute the partial decay width of $T^a_{c\bar{s}0}(2900)^{++}\to{}K^{+}D^{+}$ reaction. The corresponding Feynman diagrams
are shown in Fig.~(\ref{cc3}).  To compute the diagrams, we require the effective Lagrangian densities for the relevant interaction vertices.  Since
$T^a_{c\bar{s}0}(2900)^{++}$ resonance can be identified as an $S$-wave $D^{*}K^{*}-\rho{}D_s^{*}$ molecule~\cite{Duan:2023lcj}, the Lagrangian densities for
$T^a_{c\bar{s}0}(2900)^{++}{}D^{*}K^{*}$ and $T^a_{c\bar{s}0}(2900)^{++}{}\rho{}D_s^{*}$ vertices can be written down as~\cite{Duan:2023qsg}
\begin{align}
{\cal{L}}_{T^{a}_{c\bar{s}0}K^{*}D^{*}}&=g_{T^{a}_{c\bar{s}0}K^{*}D^{*}}D^{*\mu}\vec{\tau}\cdot\vec{T}^{a}_{c\bar{s}0}K^{*}_{\mu},\\
{\cal{L}}_{T^{a}_{c\bar{s}0}\rho{}D_s^{*}}&=g_{T^{a}_{c\bar{s}0}\rho{}D_s^{*}}D_s^{*\mu}\vec{\rho}_{\mu}\cdot\vec{T}^{a}_{c\bar{s}0},
\end{align}
where the coupling constants  $g_{T^{a}_{c\bar{s}0}K^{*}D^{*}}=2.198-5.531$ GeV and $g_{T^{a}_{c\bar{s}0}\rho{}D_s^{*}}=2.082-5.379$ GeV, which
correspond to the physical sheet~\cite{Duan:2023lcj}.

Considering the heavy quark limit and chiral symmetry, the relevant phenomenological Lagrangians
for ${\cal{D}}^{*}{\cal{D}}{\cal{P}}$ and ${\cal{D}}^{*}{\cal{D}}{\cal{V}}$ vertices are~\cite{Yue:2022mnf,Casalbuoni:1996pg,Altenbuchinger:2013vwa}
\begin{align}
{\cal{L}}_{{\cal{D}}^{*}{\cal{D}}{\cal{P}}}&=ig\langle{\cal{D}}^{*}_{\mu}u^{\mu}{\cal{D}}^{\dagger}-{\cal{D}}u^{\mu}{\cal{D}}^{*\dagger}_{\mu}\rangle,\\
{\cal{L}}_{{\cal{D}}^{*}{\cal{D}}{\cal{V}}}&=-2f_{D^{*}D{\cal{V}}}\epsilon_{\mu\nu\alpha\beta}(\partial^{\mu}{\cal{V}}^{\nu})^{i}_{j}\nonumber\\
                                           &\times({\cal{D}}_i \overleftrightarrow{\partial}_{\alpha}{\cal{D}}^{*\beta{}j\dagger}-{\cal{D}}_i^{*\beta}\overleftrightarrow{\partial}_{\alpha}{\cal{D}}^{j\dagger}),
\end{align}
where the $\langle...\rangle$ denotes trace in the SU(3) flavor space and $\epsilon^{0123}=1$. ${\cal{P}}$ and ${\cal{V}}^{\mu}$ are the SU(3) pseudoscalar meson and vector meson matrices, respectively,
\begin{align}
{\cal{P}}&=
\left(
  \begin{array}{ccc}
    \frac{1}{\sqrt{2}}\pi^{0}+\frac{1}{\sqrt{6}}\eta       & \pi^{+}                                                   &  K^{+}                         \\
    \pi^{-}                                                & -\frac{1}{\sqrt{2}}\pi^{0}+\frac{1}{\sqrt{6}}\eta         &  K^0                           \\
     K^{-}                                                 & \bar{K}^{0}                                               &  -\frac{2}{\sqrt{6}}\eta       \\
  \end{array}
\right),\\
{\cal{V}}_{\mu}&=
\left(
  \begin{array}{ccc}
    \frac{1}{\sqrt{2}}(\rho^{0}+\omega) & \rho^{+}                             &  K^{*+}     \\
    \rho^{-}                            & \frac{1}{\sqrt{2}}(-\rho^{0}+\omega) &  K^{*0}     \\
     K^{*-}                             & \bar{K}^{*0}                         &  \phi       \\
  \end{array}
\right)_{\mu},
\end{align}
and the ${\cal{D}}^{(*)}=(D^{(*)0},D^{(*)+},D_s^{*(+)})$. $u^{\mu}$ is the axial vector combination of the pseudoscalar-meson fields and at the lowest order
$u^2=-\sqrt{2}\partial^{\mu}{\cal{P}}/f_0$ with $f_0=92.4$ MeV.  The coupling constants $f_{D^{*}D{\cal{V}}}=\lambda{}m_{\rho}/(\sqrt{2}f_{\pi})$ with $\lambda=0.56$ GeV$^{-1}$
,$f_{\pi}$=132 MeV~\cite{Casalbuoni:1996pg}, and the $m_{\rho}$ is the mass of the $\rho$ meson.   The coupling constant $g=1.097$ is determined from the strong decay width $\Gamma(D^{*+}\to{}D^0\pi^{+})=56.46\pm{}1.22$ keV, together with the branching ratio $BR(D^{*+}\to{}D^0\pi^{+})=(67.7\pm{}0.5)\%$.

For the $\rho{}DD$ and $KD^{*}D_s^{*}$ vertices, the following effective Lagrangian are needed~\cite{Oh:2000qr,Lin:1999ad,Azevedo:2003qh}
\begin{align}
{\cal{L}}_{DD\rho}&=ig_{DD\rho}(D^{\dagger}\vec{\tau}\cdot\vec{\rho}^{\mu}\partial_{\mu}\bar{D}-\partial_{\mu}D^{\dagger}\vec{\tau}\cdot\vec{\rho}^{\mu}\bar{D}),\\
{\cal{L}}_{KD_s^{*}D^{*}}&=-g_{KD_s^{*}D^{*}}\epsilon^{\mu\nu\alpha\beta}(\partial_{\mu}\bar{D^{*\dagger}}_{\nu}\partial_{\alpha}D_{s\beta}^{*}\bar{K}+\partial_{\mu}D^{*\dagger}_{\nu}\partial_{\alpha}\bar{D^{*}}_{s\beta}K),
\end{align}
where the coupling constant $g_{DD\rho}=2.52$~\cite{Lin:1999ad} was derived from the D-meson electric form factor in the standard framework of
the vector meson dominance model.  $g_{KD_s^{*}D^{*}}=7.0$ GeV$^{-1}$ was computed from the SU(4) relations~\cite{Azevedo:2003qh}. The charm and $K$ mesons
isodoublets were defined as
\begin{align}
\bar{D}^{(*)\dagger}&=
\left(
  \begin{array}{cc}
    \bar{D}^{(*)0} & D^{(*)-}\\
  \end{array}
\right),~~~
D^{(*)}=
\left(
  \begin{array}{cc}
    D^{(*)0}\\
    D^{(*)+} \\
  \end{array}
\right),\\
\bar{K}^{(*)\dagger}&=
\left(
  \begin{array}{cc}
    K^{(*)-} & \bar{K}^{(*)0}\\
  \end{array}
\right),~~~
K^{(*)}=
\left(
  \begin{array}{cc}
    K^{(*)+}\\
    K^{(*)0} \\
  \end{array}
\right),
\end{align}

In addition to the vertices described above, we also need the following effective
Lagrangians~\cite{Yue:2022mnf}
\begin{align}
{\cal{L}}_{{\cal{P}}{\cal{P}}{\cal{V}}}&=-ig_h\langle[{\cal{P}},\partial^{\mu}{\cal{P}}]{\cal{V}}_{\mu}\rangle,\label{eq33}\\
{\cal{L}}_{K^{*}K{\cal{V}}^{'}}&=-g_{K^{*}K{\cal{V}}^{'}}\epsilon^{\mu\nu\alpha\beta}\partial_{\alpha}\bar{K}^{*}_{\beta}\partial_{\mu}{\cal{V}}_{\nu}^{'}K+H.c.,
\end{align}
where the ${\cal{V}}^{'}$ is meson matrices,
\begin{align}
{\cal{V}}^{'}_{\mu}&=
\left(
  \begin{array}{ccc}
    \frac{1}{\sqrt{2}}(\rho^{0}+\omega) & \rho^{+}                              \\
    \rho^{-}                            & \frac{1}{\sqrt{2}}(-\rho^{0}+\omega)  \\
  \end{array}
\right)_{\mu}.
\end{align}
The coupling constants $g_{K^{*}K{\cal{V}}^{'}}=3g_h^2/(64\pi^2f_{\pi})$ with the $g_h$ is determined via measured width of $K^{*}\to\pi{}K$.
With the help of Eq.~(\ref{eq33}), the two-body decay width $K^{*}\to{}K\pi$ is related to $g_h$ as
\begin{align}
\Gamma(K^{*+}\to{}K^{0}\pi^{+})=\frac{g^2}{24\pi{}m^2_{K^{*+}}}{\cal{P}}^3_{\pi{}K^{*}}=\frac{2}{3}\Gamma_{K^{*+}},
\end{align}
where ${\cal{P}}_{\pi{}K^{*}}$ is the three-momentum of the $\pi$ in the rest frame of the  $K^{*}$.  Using the experimental strong decay width
($\Gamma_{K^{*+}}=50.3\pm{}0.8$ MeV) and the masses of the particles shown in Ref.~\cite{ParticleDataGroup:2022pth} we obtain $g_h=9.11$.

Putting all the pieces together, we obtain the following strong decay amplitudes,
\begin{align}
{\cal{M}}_a^{\pi^0}&=-i\frac{gg_hg_{T^{a}_{c\bar{s}0}}}{f_0}\int\frac{d^4q}{(2\pi)^4}q_{\mu}\frac{-g^{\mu\nu}+q_1^{\mu}q_1^{\nu}/m^2_{D^{*+}}}{q_1^2-m^2_{D^{*+}}}\nonumber\\
                   &\times{}\frac{-g^{\nu\sigma}+q_2^{\nu}q_2^{\sigma}/m^2_{K^{*+}}}{q_2^2-m^2_{K^{*+}}}(q_{\sigma}+p_{2\sigma})\frac{1}{q^2-m^2_{\pi^0}},\\
{\cal{M}}_a^{\eta}&=i\frac{gg_hg_{T^{a}_{c\bar{s}0}}}{3f_0}\int\frac{d^4q}{(2\pi)^4}q_{\mu}\frac{-g^{\mu\nu}+q_1^{\mu}q_1^{\nu}/m^2_{D^{*+}}}{q_1^2-m^2_{D^{*+}}}\nonumber\\
                   &\times{}\frac{-g^{\nu\sigma}+q_2^{\nu}q_2^{\sigma}/m^2_{K^{*+}}}{q_2^2-m^2_{K^{*+}}}(q_{\sigma}+p_{2\sigma})\frac{1}{q^2-m^2_{\eta}},\\
{\cal{M}}_b^{\rho^0}&=-i\sqrt{2}f_{D^{*}D\rho}g_{T^{a}_{c\bar{s}0}}g_{K^{*}K\rho}\int\frac{d^4q}{(2\pi)^4}\epsilon_{\mu\nu\alpha\beta}q^{\mu}\nonumber\\
                    &\times(p^{\alpha}_{1}+q_1^{\alpha})\frac{-g^{\beta\sigma}+q_1^{\beta}q_1^{\sigma}/m^2_{D^{*+}}}{q_1^2-m^2_{D^{*+}}}\frac{-g^{\sigma\eta}+q_2^{\sigma}q_2^{\eta}/m^2_{K^{*+}}}{q_2^2-m^2_{K^{*+}}}\nonumber\\
                    &\times\epsilon_{\tau\lambda\kappa\eta}q_2^{\kappa}q^{\tau}\frac{-g^{\lambda\nu}+q^{\lambda}q^{\nu}/m^2_{\rho^0}}{q^2-m^2_{\rho^0}},\\
{\cal{M}}_b^{\omega}&=i\sqrt{2}f_{D^{*}D\omega}g_{T^{a}_{c\bar{s}0}}g_{K^{*}K\omega}\int\frac{d^4q}{(2\pi)^4}\epsilon_{\mu\nu\alpha\beta}q^{\mu}\nonumber\\
                    &\times(p^{\alpha}_{1}+q_1^{\alpha})\frac{-g^{\beta\sigma}+q_1^{\beta}q_1^{\sigma}/m^2_{D^{*+}}}{q_1^2-m^2_{D^{*+}}}\frac{-g^{\sigma\eta}+q_2^{\sigma}q_2^{\eta}/m^2_{K^{*+}}}{q_2^2-m^2_{K^{*+}}}\nonumber\\
                    &\times\epsilon_{\tau\lambda\kappa\eta}q_2^{\kappa}q^{\tau}\frac{-g^{\lambda\nu}+q^{\lambda}q^{\nu}/m^2_{\omega}}{q^2-m^2_{\omega}},\\
{\cal{M}}_c^{D^0}&=-i\frac{2gg_{DD\rho}g_{T^{a}_{c\bar{s}0}\rho{}D_s^{*}}}{f_0}\int\frac{d^4q}{(2\pi)^4}p_2^{\mu}\frac{-g^{\mu\nu}+q_1^{\mu}q_1^{\nu}/m^2_{D_s^{*+}}}{q_1^2-m^2_{D_{s}^{*+}}}\nonumber\\
                   &\times{}\frac{-g^{\nu\sigma}+q_2^{\nu}q_2^{\sigma}/m^2_{\rho^{+}}}{q_2^2-m^2_{\rho^{+}}}(q_{\sigma}+p_{1\sigma})\frac{1}{q^2-m^2_{D^0}},\\
{\cal{M}}_c^{D^{*0}}&=i2g_{KD_s^{*}D^{*}}g_{D^{*}D\rho}g_{T^{a}_{c\bar{s}0}\rho{}D_s^{*}}\int\frac{d^4q}{(2\pi)^4}\epsilon_{\mu\nu\alpha\beta}q^{\mu}q_1^{\alpha}\nonumber\\
                    &\times\frac{-g^{\beta\sigma}+q_1^{\beta}q_1^{\sigma}/m^2_{D_s^{*+}}}{q_1^2-m^2_{D_{s}^{*+}}}\frac{-g^{\sigma\eta}+q_2^{\sigma}q_2^{\eta}/m^2_{\rho^{+}}}{q_2^2-m^2_{\rho^{+}}}\nonumber\\
                   &\times{}\epsilon_{\tau\eta\lambda\kappa}q_2^{\tau}(q^{\lambda}+p_{1}^{\lambda})\frac{-g^{\nu\kappa}+q^{\nu}q^{\kappa}/m^2_{D^{*0}}}{q^2-m^2_{D^{*0}}},\\
{\cal{M}}_d^{K^{0}}&=i\frac{\sqrt{2}gg_hg_{T^{a}_{c\bar{s}0}\rho{}D_s^{*}}}{f_0}\int\frac{d^4q}{(2\pi)^4}p_{1}^{\mu}\nonumber
\end{align}
\begin{align}
                     &\times\frac{-g^{\mu\nu}+q_1^{\mu}q_1^{\nu}/m^2_{D_s^{*+}}}{q_1^2-m^2_{D_{s}^{*+}}}\frac{-g^{\nu\eta}+q_2^{\nu}q_2^{\eta}/m^2_{\rho^{+}}}{q_2^2-m^2_{\rho^{+}}}\nonumber\\
                     &\times{}(p_{2\eta}+q_{\eta})\frac{1}{q^2-m^2_{K^{0}}},\\
{\cal{M}}_d^{K^{*0}}&=i2g_{K^{*}D_s^{*}D}g_{K^{*}K\rho}g_{T^{a}_{c\bar{s}0}\rho{}D_s^{*}}\int\frac{d^4q}{(2\pi)^4}\epsilon_{\mu\nu\alpha\beta}q^{\mu}(q_1^{\alpha}+p_1^{\alpha})\nonumber\\
                     &\times\frac{-g^{\beta\sigma}+q_1^{\beta}q_1^{\sigma}/m^2_{D_s^{*+}}}{q_1^2-m^2_{D_{s}^{*+}}}\frac{-g^{\sigma\eta}+q_2^{\sigma}q_2^{\eta}/m^2_{\rho^{+}}}{q_2^2-m^2_{\rho^{+}}}\nonumber\\
                    &\times{}\epsilon_{\tau\eta\lambda\kappa}q_2^{\tau}q^{\lambda}\frac{-g^{\nu\kappa}+q^{\nu}q^{\kappa}/m^2_{K^{*0}}}{q^2-m^2_{K^{*0}}},
\end{align}
where $m_{D_s^{*}}$, $m_{D}$, $M_{D^{*}}$, $m_{K^{*}}$, and $m_{K}$ are the masses of the $D_s^{*}$, $D$, $D^{*}$, $K^{*}$, and $K$ mesons, respectively.
It is evident that these amplitudes suffer from ultraviolet (UV) divergence. Nevertheless, even the loops that are UV finite receive contributions from short distances when integrated over the entire momentum space. To address this, we will utilize a UV regulator, as described in~\cite{Lin:2017mtz}, which effectively suppresses the short-distance contributions, thereby rendering the amplitudes UV finite. The UV regulator takes the form
 \begin{align}
 \tilde{\Phi}(p_E^2/\Lambda^2)\equiv\exp{(-p_E^2/\Lambda^2)},
 \end{align}
 where $P_E$ is the Euclidean Jacobi momentum defined as $P_E=m_{i}p_j/(m_{i}+m_j)-m_jp_i(m_i+m_j)$ for the $(ij)$ molecule.

Furthermore, we adopt the dipole form factor ${\cal{F}}(q^2)=(\Lambda^2-m^2)/(\Lambda^2-q^2)$ to account for the off-shell effect of the exchanged mesons. In this expression, $m$ and $q$ represent the mass and four-momentum of the exchanged mesons, respectively. The parameter $\Lambda$ is typically parameterized as $\Lambda=m+\alpha\Lambda_{QCD}$, where $\Lambda_{QCD}=220$ MeV. The value of $\alpha$ is chosen to be approximately 1.0 to ensure that $\Lambda$ closely aligns with the mass of the exchanged mesons. In this study, we consider a range of $\alpha$ values within $0.91 \leq \alpha \leq 1.0$, a range derived from experimental data~\cite{Dong:2017rmg}.  Then we have
\begin{align}
{\cal{M}}_{total}=\sum_{i=a,b,c,d}{\cal{M}}_i\tilde{\Phi}(p_E^2/\Lambda^2){\cal{F}}^2(q^2).
\end{align}

Once the amplitudes are determined, the corresponding partial decay widths can be obtained, which read,
\begin{align}
\Gamma(T^{a}_{c\bar{s}0}(2900)^{++}\to{}K^{+}D^{+})=\frac{1}{8\pi}\frac{|\vec{p}_{K^{+}}|}{m^2_{T^{a}_{c\bar{s}0}(2900)^{++}}}\overline{|{\cal{M}}|^2},
\end{align}
where the $\vec{p}_{K^{+}}$ is the three-momenta of the decay products in the center of mass frame, the overline indicates the sum over
the polarization vectors of the final hadrons.


%


\begin{thebibliography}{23}%
\bibitem{Godfrey:1985xj}
S.~Godfrey and N.~Isgur,
Phys. Rev. D \textbf{32}, 189-231 (1985).


\bibitem{Koniuk:1979vy}
R.~Koniuk and N.~Isgur,
Phys. Rev. D \textbf{21}, 1868 (1980)
[erratum: Phys. Rev. D \textbf{23}, 818 (1981)].


\bibitem{Isgur:1978wd}
N.~Isgur and G.~Karl,
Phys. Rev. D \textbf{19}, 2653 (1979)
[erratum: Phys. Rev. D \textbf{23}, 817 (1981)].



\bibitem{Zou:2005xy}
B.~S.~Zou and D.~O.~Riska,
Phys. Rev. Lett. \textbf{95}, 072001 (2005).

\bibitem{Riska:2005bh}
D.~O.~Riska and B.~S.~Zou,
Phys. Lett. B \textbf{636}, 265-269 (2006).


\bibitem{An:2005cj}
C.~S.~An, D.~O.~Riska and B.~S.~Zou,
Phys. Rev. C \textbf{73}, 035207 (2006).


\bibitem{Zou:2010tc}
B.~S.~Zou,
Nucl. Phys. A \textbf{835}, 199-206 (2010).


\bibitem{Zou:2009zz}
B.~S.~Zou,
Nucl. Phys. A \textbf{827}, 333C-335C (2009).


\bibitem{Liu:2005pm}
B.~C.~Liu and B.~S.~Zou,
Phys. Rev. Lett. \textbf{96}, 042002 (2006).


\bibitem{Xie:2007qt}
J.~J.~Xie, B.~S.~Zou and H.~C.~Chiang,
Phys. Rev. C \textbf{77}, 015206 (2008).



\bibitem{Belle:2003nnu}
S.~K.~Choi \textit{et al.} [Belle],
Phys. Rev. Lett. \textbf{91}, 262001 (2003).



\bibitem{LHCb:2015yax}
R.~Aaij \textit{et al.} [LHCb],
Phys. Rev. Lett. \textbf{115}, 072001 (2015).


\bibitem{LHCb:2016ztz}
R.~Aaij \textit{et al.} [LHCb],
Phys. Rev. Lett. \textbf{117}, 082002 (2016).


\bibitem{LHCb:2016lve}
R.~Aaij \textit{et al.} [LHCb],
Phys. Rev. Lett. \textbf{117}, 082003 (2016).


\bibitem{LHCb:2019kea}
R.~Aaij \textit{et al.} [LHCb],
Phys. Rev. Lett. \textbf{122}, 222001 (2019).


\bibitem{LHCb:2020jpq}
R.~Aaij \textit{et al.} [LHCb],
Sci. Bull. \textbf{66}, 1278-1287 (2021).



\bibitem{LHCb:2022ogu}
R.~Aaij \textit{et al.} [LHCb],
Phys. Rev. Lett. \textbf{131},031901 (2023).



\bibitem{LHCb:2022sfr}
R.~Aaij \textit{et al.} [LHCb],
Phys. Rev. Lett. \textbf{131}, 041902 (2023).



\bibitem{Chen:2022svh}
R.~Chen and Q.~Huang,
[arXiv:2208.10196 [hep-ph]].


\bibitem{Yue:2022mnf}
Z.~L.~Yue, C.~J.~Xiao and D.~Y.~Chen,
Phys. Rev. D \textbf{107}, 034018 (2023).



\bibitem{Agaev:2022eyk}
S.~S.~Agaev, K.~Azizi and H.~Sundu,
Phys. Rev. D \textbf{107}, 094019 (2023).


\bibitem{Duan:2023lcj}
M.~Y.~Duan, M.~L.~Du, Z.~H.~Guo, E.~Wang and D.~Y.~Chen,
[arXiv:2307.04092 [hep-ph]].




\bibitem{Ke:2022ocs}
H.~W.~Ke, Y.~F.~Shi, X.~H.~Liu and X.~Q.~Li,
Phys. Rev. D \textbf{106},114032 (2022).






\bibitem{Liu:2022hbk}
F.~X.~Liu, R.~H.~Ni, X.~H.~Zhong and Q.~Zhao,
Phys. Rev. D \textbf{107}, 096020 (2023).


\bibitem{Wei:2022wtr}
J.~Wei, Y.~H.~Wang, C.~S.~An and C.~R.~Deng,
Phys. Rev. D \textbf{106}, 096023 (2022).


\bibitem{Yang:2023evp}
X.~S.~Yang, Q.~Xin and Z.~G.~Wang,
Int. J. Mod. Phys. A \textbf{38}, 2350056 (2023).


\bibitem{Lian:2023cgs}
D.~K.~Lian, W.~Chen, H.~X.~Chen, L.~Y.~Dai and T.~G.~Steele,
[arXiv:2302.01167 [hep-ph]].




\bibitem{Jiang:2023rcn}
C.~Jiang, Y.~Jin, S.~Y.~Li, Y.~R.~Liu and Z.~G.~Si,
Symmetry \textbf{15}, 695 (2023).


\bibitem{Dmitrasinovic:2023eei}
V.~Dmitra\v{s}inovi\'c,
[arXiv:2301.05471 [hep-ph]].



\bibitem{Ortega:2023azl}
P.~G.~Ortega, D.~R.~Entem, F.~Fernandez and J.~Segovia,
[arXiv:2305.14430 [hep-ph]].



\bibitem{Molina:2022jcd}
R.~Molina and E.~Oset,
Phys. Rev. D \textbf{107},056015 (2023).


\bibitem{Ge:2022dsp}
Y.~H.~Ge, X.~H.~Liu and H.~W.~Ke,
Eur. Phys. J. C \textbf{82},955 (2022).


\bibitem{Obraztsov:2016lhp}
V.~Obraztsov [OKA],
Nucl. Part. Phys. Proc. \textbf{273-275}, 1330-1333 (2016).


\bibitem{Velghe:2016jjw}
B.~Velghe [NA62-RK and NA48/2],
Nucl. Part. Phys. Proc. \textbf{273-275}, 2720-2722 (2016).


\bibitem{Quintans:2022utc}
C.~Quintans [AMBER],
Few Body Syst. \textbf{63}, 72 (2022).




\bibitem{Nagae:2008zz}
T.~Nagae,
Nucl. Phys. A \textbf{805}, 486-493 (2008).



\bibitem{Dong:2014ksa}
Y.~Dong, A.~Faessler, T.~Gutsche and V.~E.~Lyubovitskij,
Phys. Rev. D \textbf{90}, 094001 (2014).



\bibitem{Zhu:2019vnr}
H.~Zhu and Y.~Huang,
Phys. Rev. D \textbf{100},054031 (2019)..




\bibitem{Dong:2010xv}
Y.~Dong, A.~Faessler, T.~Gutsche, S.~Kumano and V.~E.~Lyubovitskij,
Phys. Rev. D \textbf{82}, 034035 (2010).



\bibitem{Duan:2023qsg}
M.~Y.~Duan, E.~Wang and D.~Y.~Chen,
[arXiv:2305.09436 [hep-ph]].


\bibitem{ParticleDataGroup:2022pth}
R.~L.~Workman \textit{et al.} [Particle Data Group],
PTEP \textbf{2022}, 083C01 (2022).



\bibitem{Zhu:2019vnr}
H.~Zhu and Y.~Huang,
Phys. Rev. D \textbf{100}, 054031 (2019).



\bibitem{He:2011jp}
J.~He, Z.~Ouyang, X.~Liu and X.~Q.~Li,
Phys. Rev. D \textbf{84}, 114010 (2011).



\bibitem{Xie:2015zga}
J.~J.~Xie, Y.~B.~Dong and X.~Cao,
Phys. Rev. D \textbf{92},034029 (2015).



\bibitem{Lebiedowicz:2011tp}
P.~Lebiedowicz and A.~Szczurek,
Phys. Rev. D \textbf{85}, 014026 (2012).


\bibitem{Lebiedowicz:2011nb}
P.~Lebiedowicz, R.~Pasechnik and A.~Szczurek,
Phys. Lett. B \textbf{701}, 434-444 (2011).


\bibitem{Belle:2007qxm}
G.~Pakhlova \textit{et al.} [Belle],
Phys. Rev. D \textbf{77}, 011103 (2008).


\bibitem{BaBar:2006qlj}
B.~Aubert \textit{et al.} [BaBar],
Phys. Rev. D \textbf{76}, 111105 (2007).


\bibitem{Guo:2016iej}
X.~D.~Guo, D.~Y.~Chen, H.~W.~Ke, X.~Liu and X.~Q.~Li,
Phys. Rev. D \textbf{93},054009 (2016).











\bibitem{Casalbuoni:1996pg}
R.~Casalbuoni, A.~Deandrea, N.~Di Bartolomeo, R.~Gatto, F.~Feruglio and G.~Nardulli,
Phys. Rept. \textbf{281}, 145-238 (1997).


\bibitem{Altenbuchinger:2013vwa}
M.~Altenbuchinger, L.~S.~Geng and W.~Weise,
Phys. Rev. D \textbf{89}, no.1, 014026 (2014).


\bibitem{Oh:2000qr}
Y.~s.~Oh, T.~Song and S.~H.~Lee,
Phys. Rev. C \textbf{63}, 034901 (2001).


\bibitem{Lin:1999ad}
Z.~w.~Lin and C.~M.~Ko,
Phys. Rev. C \textbf{62}, 034903 (2000).


\bibitem{Azevedo:2003qh}
R.~S.~Azevedo and M.~Nielsen,
Phys. Rev. C \textbf{69}, 035201 (2004).


\bibitem{Lin:2017mtz}
Y.~H.~Lin, C.~W.~Shen, F.~K.~Guo and B.~S.~Zou,
Phys. Rev. D \textbf{95},114017 (2017).


\bibitem{Dong:2017rmg}
Y.~Dong, A.~Faessler, T.~Gutsche, Q.~L\"u and V.~E.~Lyubovitskij,
Phys. Rev. D \textbf{96},074027 (2017).



\bibitem{Kim:2017nxg}
S.~H.~Kim, S.~i.~Nam, D.~Jido and H.~C.~Kim,
Phys. Rev. D \textbf{96},014003 (2017).


\end{thebibliography}
\end{document}